\documentclass[11pt]{article}
\usepackage[utf8]{inputenc}

\usepackage{comment}

\usepackage[ruled,linesnumbered]{algorithm2e}
\usepackage{xcolor}
\usepackage{graphicx,natbib,amsfonts,amsthm,amssymb}
\usepackage[colorlinks,citecolor=blue,urlcolor=blue]{hyperref}
\usepackage{relsize}
\usepackage{amsmath,mathrsfs}
\usepackage{wrapfig,lipsum,booktabs}
\usepackage{subcaption}
\usepackage{appendix}
\usepackage{enumitem}
\usepackage{multirow}
\usepackage{optidef}

\usepackage{xr}

\newcommand{\blind}{1}

\def\spacingset#1{\renewcommand{\baselinestretch}%
{#1}\small\normalsize} \spacingset{1}

\newcommand{\IP}{{\rm I}\kern-0.18em{\rm P}}
\newcommand{\1}{{\rm 1}\kern-0.24em{\rm I}}
\newcommand{\E}{{\rm I}\kern-0.18em{\rm E}}
\newcommand{\R}{{\rm I}\kern-0.18em{\rm R}}

\newcommand{\cS}{\mathcal{S}}

\newcommand{\qmq}[1]{\quad\mbox{#1}\quad}

\newcommand{\tO}{\Tilde{O}}
\newcommand{\tOmega}{\Tilde{\Omega}}

\newcommand{\stalpha}{\alpha^\star}

\newcommand{\stbeta}{\beta^\star}
\newcommand{\stZ}{Z^\star}
\newcommand{\stz}{z^\star}
\newcommand{\cstZ}{\mathcal{Z^\star}}
\newcommand{\cU}{\mathcal{U}}
\newcommand{\stTheta}{\Theta^\star}
\newcommand{\stG}{G^\star}

\newcommand{\F}{\mathrm{F}}

\newcommand{\cF}{\mathcal{F}}
\newcommand{\cE}{\mathcal{E}}
\newcommand{\cC}{\mathcal{C}}
\newcommand{\cA}{\mathcal{A}}

\newcommand{\cO}{\mathcal{O}}

\newcommand{\hG}{{\widehat{G}}}
\newcommand{\halpha}{{\widehat{\alpha}}}
\newcommand{\hbeta}{{\widehat{\beta}}}
\newcommand{\hTheta}{{\widehat{\Theta}}}

\newcommand{\op}{\mathrm{op}}

\newtheorem{theorem}{Theorem}
\newtheorem{example}{Example}
\newtheorem{assumption}{Assumption}
\newtheorem{remark}{Remark}
\newtheorem{lemma}{Lemma}

\newtheorem{Corollary}{Corollary}

\newtheorem{setting}{Setting}

\theoremstyle{definition}
\newtheorem{simulation}{Simulation}

\renewcommand{\appendix}{
 \setcounter{section}{0}%
  \setcounter{subsection}{0}%
  \renewcommand\thesection{\Alph{section}}
  \setcounter{equation}{0}
  \renewcommand{\theequation}{S.\arabic{equation}}
  \setcounter{figure}{0}
  \renewcommand\thefigure{S\arabic{figure}}
  \setcounter{table}{0}
  \renewcommand\thetable{S\arabic{table}}  
    \setcounter{lemma}{0}
  \renewcommand\thelemma{S\arabic{lemma}}  
    \setcounter{proposition}{0}
  \renewcommand\theproposition{S\arabic{proposition}}  
    \setcounter{setting}{0}
  \renewcommand\thesetting{S\arabic{setting}}  
    \setcounter{Corollary}{0}
  \renewcommand\theCorollary{S\arabic{Corollary}}  
    \setcounter{Definition}{0}
  \renewcommand\theDefinition{S\arabic{Definition}}  
    \setcounter{theorem}{0}
  \renewcommand\thetheorem{S\arabic{theorem}}  
      \setcounter{example}{0}
  \renewcommand\theexample{S\arabic{example}}
   \setcounter{remark}{0}
  \renewcommand\theremark{S\arabic{remark}}  
  \setcounter{algocf}{0}
   \renewcommand\thealgocf{S\arabic{algocf}} 
  \setcounter{simulation}{0}
   \renewcommand\thesimulation{S\arabic{simulation}} 
  }

\allowdisplaybreaks

\begin{document}

\if1\blind
{
  \title{Local Information for Global Network Estimation in Latent Space Models}
  \author{Lijia Wang\\
    Department of Data Science, City University of Hong Kong\\
    Xiao Han\\
     International Institute of Finance, School of Management,\\ University of Science and Technology of China\\
Yanhui Wu   \\
Faculty of Business and Economics, University of Hong Kong\\
    Y. X. Rachel Wang\hspace{.1cm} \thanks{Correspondence should be addressed to Y.X. Rachel Wang (rachel.wang@sydney.edu.au)} \hspace{.2cm}\\
    School of Mathematics and Statistics, University of Sydney
}
  \maketitle
} \fi

\if0\blind
{
	\bigskip
	\bigskip
	\bigskip
	\begin{center}
		{\LARGE\bf Local Information for Global Network Estimation in Latent Space Models}
	\end{center}
	\medskip
} \fi

\def\spacingset#1{\renewcommand{\baselinestretch}%
{#1}\small\normalsize} \spacingset{1}

\begin{abstract}
In many social networks, an individual observes only a restricted local view of the full network structure. We study such local views under a partial information framework that models an individual’s observations as a subgraph based on path length, and address the problem of estimating a general latent space model from a single individual’s local view. Compared to the full network, the partial information network contains many missing edges and depends on a random, potentially sparse neighborhood, posing significant challenges for estimation. We propose a projected gradient descent algorithm for maximum likelihood estimation and establish theoretical guarantees for its convergence under both conditional likelihood and full likelihood settings. To characterize the quality of a local view, we introduce an imbalance measure as a theoretical and diagnostic quantity for assessing bias in a local view and show that it plays a central role in determining convergence rates and estimation error bounds. Using simulated networks, we demonstrate that satisfactory estimation is possible from a single local view. In an application to U.S. Congress co‑sponsorship networks, we show how the estimated latent positions reveal nuanced structure in legislators’ social relationships.
\end{abstract}

\noindent%

\vfill

\newpage

\spacingset{1.9}

\section{Introduction }

Network data arise across disciplines \citep{newman2018networks} and encode complex dependence through relationships among subjects. A large statistical literature studies network inference, including community detection  \citep{amini2013pseudo, lei2015consistency,wang2023fast},  community
structure testing \citep{yuan2022likelihood,jin2021optimal}, membership profiling \citep{fan2022simple}, and latent position estimation \citep{hoff2002latent,ma2020universal,tang2024population, he2024semiparametricmodelinganalysislongitudinal}, typically assuming that the full network is observed.

Another line of work infers global properties by aggregating multiple subgraphs, motivated by hard-to-reach populations and divide-and-conquer computation at scale. In the network sampling literature, methods such as egocentric \citep{freeman1982centered,wasserman1994social}, snowball \citep{goodman1961snowball}, and respondent-driven sampling \citep{heckathorn1997respondent} first select nodes as ``seeds" and then expand into their neighborhoods using suitable criteria. An example of such a subgraph is an ego network (egonet) \citep{crossley2015social,sengupta2018anomaly}. These approaches aggregate node-specific information across many local neighborhoods to estimate population-level attributes (e.g., disease prevalence, behavior traits), and under specific model assumptions can estimate model parameters \citep{handcock2010modeling} and community structure \citep{mukherjee2021two,chakrabarty2025subsampling}.

Our framework instead studies the network structure visible to a \textit{single} individual and how the information content of such locally observed structures varies across nodes. In contrast, most existing work either aggregates information from multiple subgraphs or assumes access to sufficiently rich samples to support global estimation, without explicitly quantifying what can be learned from an individual’s local view.
This distinction is important because, in practice, individuals in a social network often only have access to localized knowledge within their neighborhoods. In economics and social science, such individual-level network knowledge is essential for understanding individual behaviors and studying peer effects \citep{jackson2014culture,troost2023neighbourhood}.  
In this paper, we focus on the problem of estimating global network parameters using the local network around a given individual and examine the influence of neighborhood structure on estimation accuracy.

\subsection{Related work and preview of contributions}\label{sec:preliminary}

To characterize an individual’s local network, we adopt the individual-centered partial information framework proposed by \cite{han2020individual}. The framework assumes that an individual in a social network forms a local perspective about the global network by gathering information through their $L$-hop neighborhood for a prespecified path length $L$. We discuss such partial information networks and compare them with egonets, another commonly used local network representation, in Section~\ref{subsec:partial_set}.  \citet{han2020individual} studies the problem of community detection using a partial information network under the stochastic block models (SBM) \citep{holland1983stochastic, karrer2011stochastic}, which assign nodes to discrete communities. In comparison, we consider a class of general latent space models \citep{hoff2002latent}, which characterize social structure by capturing node proximity beyond discrete communities. We study the problem of estimating global model parameters using the partial information network of an individual (or node).




In the full-network setting, latent space parameter estimation is generally harder than community detection in SBMs, as it requires estimating continuous latent vectors rather than discrete labels. Accordingly, SBMs admit stronger guarantees (exact recovery \citep{abbe2015exact,Zhang2016MINIMAX}; sharp spectral perturbation bounds \citep{cape2019two}) in suitable regimes.
In comparison, results for latent space models usually bound latent position error; e.g., in random dot product graphs the minimax error rate is $n^{-1/2}$ up to polylogarithmic factors \citep{yan2023minimax}, and more generally the rates depend on latent dimension, sparsity, link-function smoothness, and embedding approximation quality
\citep{tang2013universally,ma2020universal,he2024semiparametricmodelinganalysislongitudinal, tian2026efficient}.

Given these differences in the full-network setting, we expect them to persist under the partial information framework, and thus our estimation and analysis differ substantially from those of \citet{han2020individual}.  While a partial information network can be considered as a full network with missing edges, the pattern of missing edges is notably different from conventional structures \citep{candes2010matrix, abbe2020entrywise}, making existing matrix completion methods inapplicable.  
We make the following main contributions in this paper:
\begin{itemize}[leftmargin=1.2em, itemsep=0ex, topsep=0ex]
    \item 
    We develop a maximum likelihood approach for fitting a latent space model from a given node’s partial information network, and propose an efficient projected gradient descent algorithm that estimates latent positions for all nodes, along with other estimated model parameters, for visualizing node similarities and distances.  
    \item We establish convergence and error bounds by analyzing the algorithm first under a conditional likelihood  and then generalizing to the full likelihood, thereby overcoming challenges associated with  the randomness and sparsity of observed neighborhoods.



\item We characterize how neighborhood structure affects accuracy and convergence, introducing a neighborhood imbalance measure, {primarily as a theoretical and diagnostic tool} for quantifying local-view bias; simulations and real data show this measure is strongly associated with estimation error.


\end{itemize}

\spacingset{1}
\begin{figure}
        \centering
        \hspace{-1cm}
       \begin{subfigure}[b]{0.32\textwidth}
          \includegraphics[width = \textwidth]{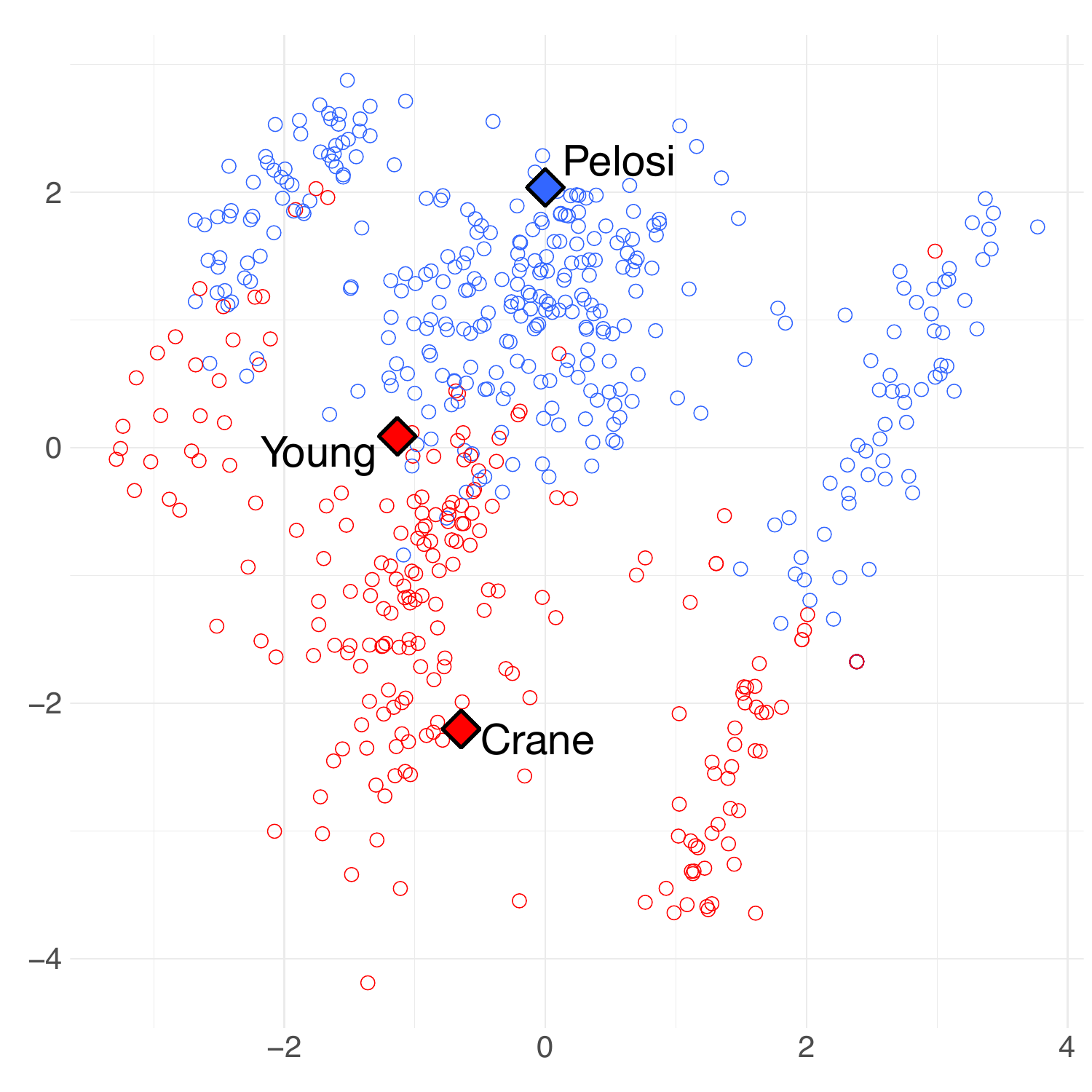}
    \caption{Entire network} \label{fig:85_90_congress_1} 
    \end{subfigure}
      \begin{subfigure}[b]{0.32\textwidth}
          \includegraphics[width = \textwidth]{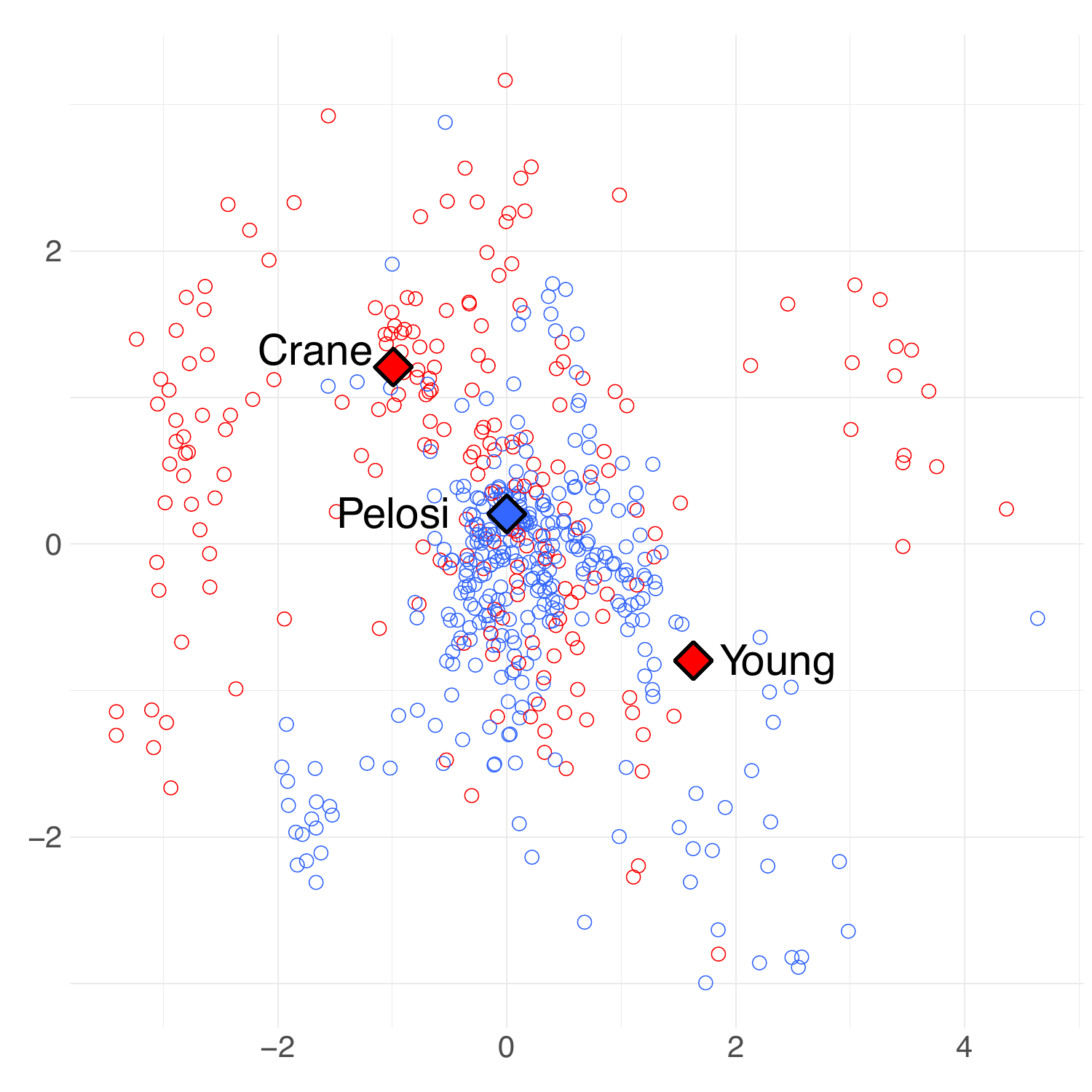}
    \caption{Armey}\label{fig:85_90_congress_2} 
    \end{subfigure}
    \begin{subfigure}[b]{0.32\textwidth}
          \includegraphics[width = \textwidth]{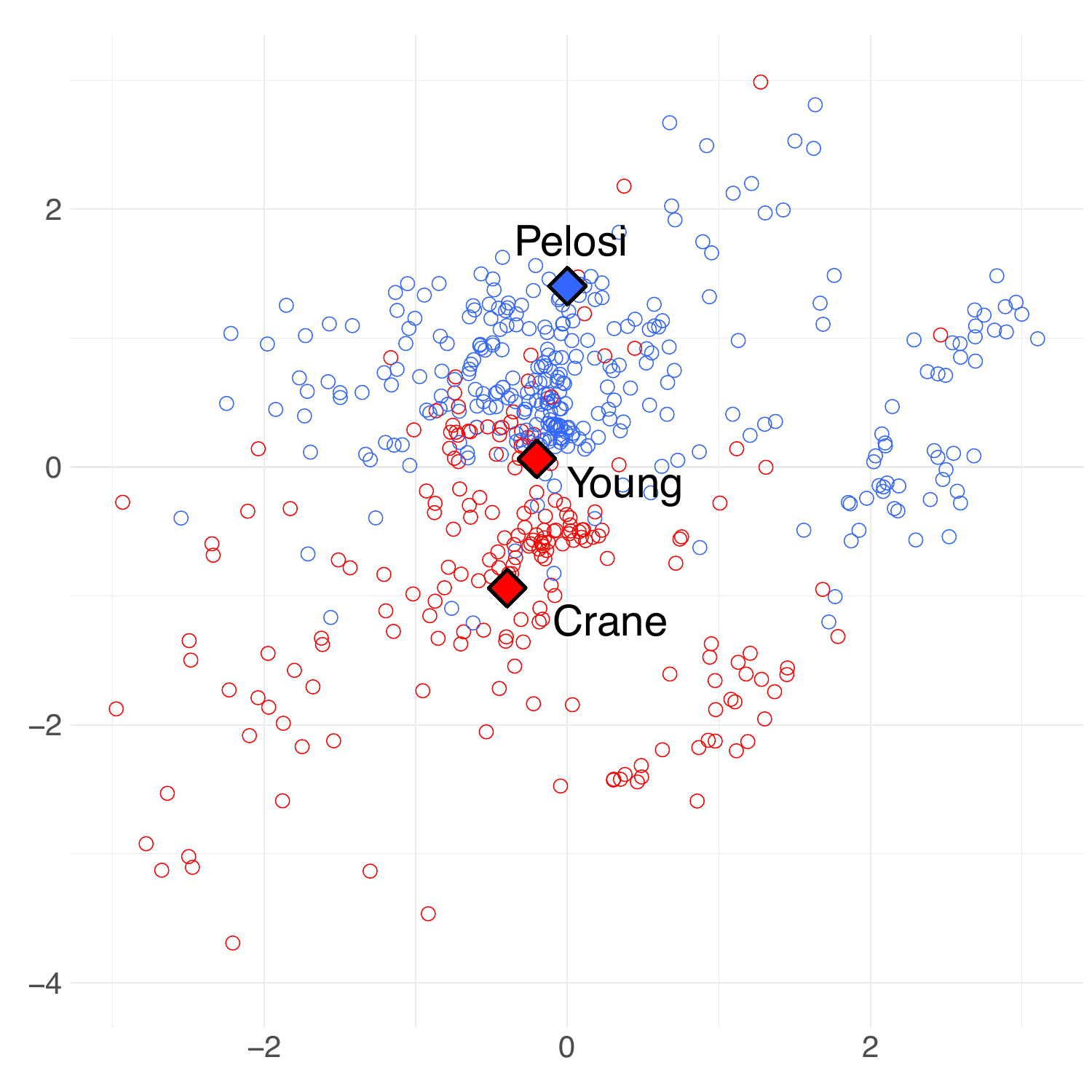}
    \caption{Furse} \label{fig:85_90_congress_3} 
    \end{subfigure}
    \caption{Latent positions of U.S. Congress legislators in 1990-1994, estimated using
    (a) the entire network; the partial information network of (b) Armey and (c) Furse.  The points are color-coded according to party affiliation. } \label{fig:85_90_congress_11} 
\end{figure}
\spacingset{1.9}



A preview of our results from the U.S. Congress co-sponsorship network \citep{fowler2006connecting,fowler2006legislative}, which represents collaborative relationships between legislators, is given in Figure~\ref{fig:85_90_congress_11}. The latent positions of the legislators in Figure~\ref{fig:85_90_congress_1} are estimated using the entire network, whereas Figures~\ref{fig:85_90_congress_2} and~\ref{fig:85_90_congress_3} use only the local networks of two house representatives, Richard Armey and Elizabeth Furse, respectively, for estimation. For comparison, we indicate the positions of Phil Crane, Bill Young, and Nancy Pelosi in all three figures as points of reference. The results from the local network of Furse align better with the results from the full network, especially for the relative positions of the three reference points. As we will explain in Section~\ref{subsec:congress}, this observation is consistent with Furse having a more balanced neighborhood despite having a similar degree as Armey. Furthermore, we observe that there is a tradeoff between degrees and neighborhood imbalance, suggesting that legislators with high degrees tend to have more biased local views. Our analysis provides a unique perspective compared to existing statistical studies of this network, many of which have focused on global community detection or latent position estimation. Instead, we characterize when an individual's local view exhibits a smaller bias and the utility of such a local view in global estimation.



\subsection{Outline of the paper}

The rest of the paper is organized as follows. Section~\ref{sec:prob_set} explains the framework of individual-centered partial information in social networks and introduces the problem setup for latent space models.
Section~\ref{sec:algorithm_setting} proposes a projected gradient descent algorithm for solving the maximum likelihood problem and estimating the parameters. Section~\ref{sec:theoretical_results} establishes convergence results for the algorithm in different settings, emphasizing the importance of neighborhood effect. The performance of our method and the influence of neighborhood bias are examined using simulated and real data in Sections~\ref{sec:simulation_study} and~\ref{sec:real_data}, respectively.
Section~\ref{sec:discussion} concludes with a discussion and ideas for future work. All technical proofs, additional algorithm and numerical results can be found in the Supplementary Materials.

\section{Problem setup}\label{sec:prob_set}

\subsection{Individual-centered partial information in a social network}
\label{subsec:partial_set}

Consider an undirected binary social network represented by a symmetric adjacency matrix $A \in \{0,1\}^{n\times n}$, where $A_{ij} = 1$ if nodes $i$ and $j$ are connected, and $A_{ij} = 0$ otherwise. Without loss of generality, assume node 1 is the individual of interest. The individual-centered partial information framework in \citet{han2020individual} introduces  ``knowledge depth'' to depict an individual's perception of the network. An individual $i$ has a knowledge depth of $L$ when they can observe all paths originating from $i$ up to length $L$ within the network. 

When $L=1$, the observed subgraph is a star around $i$ with no edges among neighbors, which is typically insufficient for global estimation. Increasing $L$ expands both node and edge visibility (i.e., proportion of observed nodes/edges), but large $L$ is often unnecessary. “Six degrees of separation” \citep{watts1998collective} implies that  small path lengths can already cover much of the network.
In the U.S. Congress co-sponsorship network (Supplementary Figure S11), $L=3$ essentially reveals the full network, whereas $L=1$ reaches only about 10\% of nodes. The intermediate case $L = 2$ strikes a balance with high node visibility but limited edge visibility (around 20\% here), making inference informative but nontrivial. In practice, 
$L=2$ also matches information constraints on many social media platforms (e.g., Instagram), where users can often see friends and friends-of-friends but privacy settings restrict access beyond that. {While the availability of friends-of-friends information is application-dependent,} $L=2$ is realistic and well motivated.
$L=3$ is rarely available in these cases. Therefore, similar to \citet{han2020individual}, we focus on $L = 2$, where individuals observe connections involving their friends and friends of friends. 

Let $B$ denote the partial adjacency matrix observed by node 1, then the entries in $B$ can be written as $B_{ij} = A_{ij}( 1 -  \1\{ A_{1i} = 0\} \1\{ A_{1j} = 0\})$ for $i,j \in [n]$, where $\1\{\cdot\}$ is an indicator function and  $[n]=\{1, \ldots, n\}$. In matrix form,
\vspace{-0.2cm}
\begin{equation}\label{eq:B_form}
B = SA + AS - SAS,  \qmq{where} S = \mbox{diag}(A_{11}, A_{12}, \ldots A_{1n}). \vspace{-0.2cm}
\end{equation}
We also define the neighborhood set as 
\vspace{-0.2cm}
\begin{equation}\label{eq:neighbor_set}
    \cS = \{i \in [n] \mid A_{1i} = 1\} \cup \{1\}\,, \vspace{-0.2cm}
\end{equation}
which consists of nodes directly connected to node 1 and node 1 itself for convenience. The size of the neighborhood and the neighborhood ratio are denoted as 
\vspace{-0.2cm}
$$n_S = |\cS| \qmq{and} r_S = \frac{n_S}{n}\,, \vspace{-0.2cm}$$ 
respectively, where $|\cdot|$ is the cardinality of a set.  Noting that $I-S$ is the diagonal matrix for the complement set $\cS^c$, we also define the size of this complement as $n_{I - S} = n - n_S$.

We note that the
$L=2$ partial information network differs from the egonet---a commonly used local subgraph structure in network analysis \citep{crossley2015social,sengupta2018anomaly}---in that it encompasses a larger graph. If the ego is node~1, then its egonet observes $\{A_{ij} \mid i, j \in \cS\}$, whereas the $L=2$ view observes a strictly larger set $\{A_{ij} \mid i \text{ or } j \in \cS\}$. 
This substantially improves node visibility, as illustrated in Supplementary Figure S11, and enables the estimation of global network properties from a single local view.

Given a single local view, the partial adjacency matrix $B$ is used to estimate the parameters in the latent-space model. However, the large proportion of missing entries, together with their highly non-random structure, poses significant technical challenges for estimating global network properties and can introduce biases stemming from the individual’s local perspective of the network.

\subsection{A general latent space model with covariates}
While \citet{han2020individual} studied partial information under SBMs, we consider a broader class of latent space models \citep{hoff2002latent} that incorporate distance structure and covariates in edge formation. Specifically, let $A$ be an $n\times n$ adjacency matrix and $X\in\R^{n\times n}$ be an observed covariate matrix. A popular class of latent space models, called inner-product models \citep{hoff2003random,hoff2005bilinear, handcock2007model}, assumes that for any $i \neq j \in [n]$,
\begin{equation}
A_{ij} = A_{ji} \overset{ind}{\sim} \mbox{Bernoulli} (P_{ij})
    \qmq{and} \mbox{logit}(P_{ij}) = \Theta_{ij} = \alpha_i +\alpha_j + \beta X_{ij} +z_i^\top  z_j \,, \label{eq:lsm}
\end{equation}
where the link function is $\mbox{logit}(x) = \log[x/(1- x)]$. By convention, $A_{ii}=0$  and $X_{ii}$ is also set to $0$. Other link functions have also been considered, including the probit \citep{hoff2007modeling} and identity \citep{young2007random} functions. Our method and theoretical guarantees extend to other link functions (e.g., probit link) under appropriate convexity and smoothness conditions (see Supplementary Section~C.1.4). In Eq~\eqref{eq:lsm}, $z_i\in\R^k$ denotes node $i$'s latent position (social space) and $\alpha_i\in\R$ captures degree heterogeneity. The coefficient $\beta$ measures the effect of the covariate $X_{ij}$ (discrete or continuous); for instance, if $X_{ij}$ indicates whether $i$ and $j$ share a political party, then $\beta>0$ implies higher connection propensity among similar nodes (homophily) \citep{mcpherson2001birds,huang2009virtually}.

In matrix form, the parameters in the inner-product model~\eqref{eq:lsm} can be expressed as
\begin{align}
      & \Theta = \alpha 1_n^\top  +1_n\alpha^\top  + \beta X + G \,, \notag\\
      \qmq{where} &  G = Z Z^\top  \mbox{ and  } J^0Z = Z  \mbox{ with }  J^0 = I_{n} - \frac{1}{n} 1_{n} 1^\top _{n}\,. \label{eq:J0_form}
\end{align}
Here we define $\alpha = ( \alpha_1, \ldots, \alpha_n)^\top \in \R^n$, $Z = (z_1, \ldots, z_n)^\top \in \R^{n\times k}$. $J^0$ is a commonly used centering matrix to ensure the identifiability of the model parameters. As a result, this condition uniquely identifies $Z$ up to an orthonormal transformation of its rows, i.e., $G = Z Z^\top $ is identifiable. Furthermore, the latent positions in the entire network are centered at $0$ with $\sum_{i \in [n]} z_{i} = 0$. We also assume $\Theta_{ij}$ is bounded such that $-M_1 \leq \Theta_{ij} \leq -M_2$ for $1\leq i \neq j \leq n$, and $|\Theta_{ii}| \leq M_1$ for $i \in [n]$, for some $M_1, M_2>0$. This implies  $\frac{1}{2} e^{-M_1} \leq P_{ij} \leq e^{-M_2}$. Thus, $M_1$ controls the conditioning of the problem and $M_2$ controls the sparsity of the network.  
$M_1$ and $M_2$ are allowed to vary with the network size $n$. The conventional setting assumes the entire network $A$ and the covariate matrix $X$ are observed, while the latent positions $Z$ and model parameters $\alpha$, $\beta$ need to be estimated from data. 

\subsection{Partial information in the latent space model}\label{sec:lsp_model_setting}

Under partial information, we only observe a partial adjacency matrix $B$ as defined in Eq~\eqref{eq:B_form} with many edges missing relative to $A$. Noting that $\IP(B_{ij} = 1 \mid i \,\mathrm{or}\, j \in \cS) = P_{ij}$ for $i,j \neq 1$, and $B_{ij} = 0$ for $i$ and $j \notin \cS$, where $\cS$ is the neighborhood set in Eq~\eqref{eq:neighbor_set}, we maximize the log-likelihood of the observed data to estimate $(Z, \alpha, \beta)$,
\begin{align}
    l(\Theta) =  & \sum_{i \in \cS\,,\, j > i }\left[ B_{ij} \Theta_{ij} + \log\left(1 - \sigma(\Theta_{ij})\right)\right]      \notag\\
    = &  \sum^n_{j = 2} \left[ B_{1j} \log \left( \sigma(\Theta_{1j} ) \right) + (1 - B_{1j} )\log\left(1 - \sigma(\Theta_{1j})\right)\right] \notag\\
     & + \sum_{i \in \cS \backslash \{1\}\,,\, j > i }\left[  B_{ij} \log \left( \sigma(\Theta_{ij} ) \right)+ (1 - B_{ij} )\log\left(1 - \sigma(\Theta_{ij})\right)\right] \label{eq:likelihood}
\end{align} 
 Here, $\sigma(x) = 1/(1 + e^{-x})$ is the sigmoid function with $P_{ij} = \sigma(\Theta_{ij})$ for $i \neq j$. We note that $l(\Theta)$ consists of two parts. The first part depends only on $B_{1j}$ for $j > 1$ (note that $B_{1j} = A_{1j}$), which are connections with the neighborhood set $\cS$, while the second part depends on the connections between $\cS$ and nodes that are two steps away from the center (node 1). Motivated by this observation, our theoretical analysis later proceeds by first conditioning on $\cS$ (Sections~\ref{sec:lsp_performance}-\ref{sec:neighbor_model}), followed by the full likelihood setting in Section~\ref{sec:random_neighbor_model}.

\subsection{Notation}\label{sec:notation}

We use the following notation throughout. Let $1_m=(1,\ldots,1)^\top\in\R^m$, $I_m$ be the $m\times m$ identity matrix, and $[c]=\{1,\ldots,c\}$, for positive integers $m,c$. For $X,Y\in\R^{m_1\times m_2}$, define $\langle X,Y\rangle\mbox{Trace}(X^\top  Y)$. The norms $\|\cdot\|_\op$, $\|\cdot\|_\F$, and $\|\cdot\|_*$ denote the operator, Frobenius, and nuclear norms; for vectors, $\|\cdot\|$ and $\|\cdot\|_\infty$ denote the $L_2$ and infinity norms. For a matrix $M$, let $\sigma_i(M)$ be its $i$-th largest singular value. We use standard order notation $O$, $o$, $\Omega$, and $O_p$. For positive sequences $a_n,b_n$, write $a_n\simeq b_n$ if $\limsup_{n\to\infty} a_n/b_n<\infty$ and $\limsup_{n\to\infty} b_n/a_n<\infty$. Constants such as $c,c_0,C_1$ may vary across lines and do not depend on $n$. Events hold with high probability (w.h.p.) if $\IP(\cE_n)\ge 1-n^{-c}$ for some $c>0$ and all sufficiently large $n$.


For matrix and vector structures under partial information, we generalize Eq~\eqref{eq:B_form} as a transformation $S(M) = SM +MS -SMS$ for any square matrix $M \in \R^{n\times n}$. $M_{S,I-S}$ is a submatrix that takes the $i$-th rows and $j$-th columns of $M$ with $i \in \cS$ and $j\notin \cS$. Similarly, we define submatrices $M_{S,S}$, $M_{I-S,S}$ and $M_{I-S,I-S}$. For a matrix $ M \in \R^{n \times m}$ with $m$ being any positive integer, let $M_S$ denote the submatrix that collects the $i$-th rows of $M$ with $ i \in \cS$; $M_{I-S}$ takes the $i$-th rows of $M$ with $i \notin \cS$. For any vector $v \in \R^n$, we construct subvector $v_S$ and $v_{I-S}$ in the same way.

\section{Parameter estimation}\label{sec:algorithm_setting}
In this section, we develop the method for estimating the parameters in $\Theta$, which take the form in Eq~\eqref{eq:J0_form}. We formulate the task as an optimization problem that minimizes the negative log-likelihood function in Eq~\eqref{eq:likelihood} with the identifiable constraints on $\Theta$ in Eq~\eqref{eq:J0_form}.

The optimization problem  is non-convex due to its quadratic form. To ensure computational efficiency and scalability, we adopt the projected gradient descent approach \citep{ma2020universal, tang2024population, he2024semiparametricmodelinganalysislongitudinal} to obtain parameter estimates.  It is important to note, however, that the presence of unobserved edges in 
$A_{I-S, I-S}$ necessitates modifications to both the optimization problem and the algorithm, making existing theoretical results inapplicable to our problem.



The unobserved edges in $A_{I-S, I-S}$ mean the amount of information available for updating the degree heterogeneity parameters $\alpha$ is significantly different for the neighbor and non-neighbor components. In our algorithm, we update $\alpha_S$ and $\alpha_{I-S}$ using $A_{S, I}$ and $A_{I-S, S}$ respectively, with different step sizes. Furthermore, we introduce a new centering matrix $J(\cS)$ depending on $\cS$ in the projection step of the algorithm, defined by
 \vspace{-0.2cm}
 
\spacingset{1}
\begin{equation}\label{eq:J_form}
J(\cS) = I_n - W_S \qmq{where} [W_S]_{ij} = \begin{cases}  1/n_S \,,\, & i,j \in \cS;\\
 1/n_{I - S}\,,\,& i,j \notin \cS;\\
0\,,\,& \mathrm{otherwise}.
\end{cases}
\end{equation}
\spacingset{1.9} Centering the positions of the neighbors and non-neighbors separately enables a key argument in our analysis to decouple the estimation errors for $\alpha$ and $Z$, allowing the theory to accommodate sparse networks with a shrinking neighborhood ratio. More specifically, compared to the original matrix $J^0$, the projection $J(\cS)$ enforces $\sum_{i \in \cS} z_{i} = 0$ and $\sum_{i \notin \cS} z_{i} = 0$, while still ensuring the projected $Z$ still satisfies $J^0 Z =Z$. 
In the following discussion, we will abbreviate $J(\cS)$ as $J$, with the dependence on $\cS$ understood from the context.

\spacingset{1}
\begin{algorithm}[htb!]
\caption{ Projected gradient descent method}\label{alg:lsp}
\SetKw{KwBy}{by}
\SetKwInOut{Input}{Input}\SetKwInOut{Output}{Output}

\SetAlgoLined

\Input{Partial adjacency matrix: $B$; covariate matrix: $X$; initial estimates: $Z^0$, $\alpha^0$; $\beta^0$; step sizes: $\eta_Z$, $\eta_{\alpha_S}$, $\eta_{\alpha_{I-S}}$ and $\eta_{\beta}$; constraint sets: $\cC_Z$, $\cC_\alpha$, $\cC_\beta$.}

$S = \mbox{diag}(1, B_{12}, \ldots, B_{1n})$

\For{$t = 0, 1, \ldots, T$}{

$\Theta^t = \alpha^t 1_n^\top  +1_n(\alpha^t)^\top  + \beta^t X + Z^t(Z^t)^\top $, $P^t = \sigma(\Theta^t)$

$Z^{t + 1} = Z^t +2 \eta_Z ( B -  S(P^t))Z^t$ 

$\alpha^{t + 1}_S = \alpha^t_S + 2\eta_{\alpha_S}( B_{S,I}  -  P^t_{S,I}) 1_{n}$, $\alpha^{t + 1}_{I-S} = \alpha^t_{I-S} + 2\eta_{\alpha_{I-S}}( B_{I-S,S} -  P^t_{I-S,S}) 1_{n_S}$

$\beta^{t +1} = \beta^t + \eta_\beta \langle B -  S(P^t),\, X\rangle$

$Z^{t + 1} = \mathcal{P}_{\cC_Z}(Z^{t+1})$, $\alpha^{t + 1} = \mathcal{P}_{\cC_\alpha}(\alpha^{t+1})$, $\beta^{t + 1} = \mathcal{P}_{\cC_\beta}(\beta^{t+1})$  \tcp*{
$\mathcal{P}$ denotes projection onto suitable constraint sets $\cC_Z, \cC_{\alpha}, \cC_{\beta}$.}

\label{eq:projection_alg}

}

\Output{$Z^T, \alpha^T, \beta^T$}
\end{algorithm}
\spacingset{1.9}

The full projected gradient descent algorithm is provided in Algorithm~\ref{alg:lsp}. The algorithm iteratively updates the estimates for the three parameters, $Z$, $\alpha$, and $\beta$, starting from some initial values $Z^0$, $\alpha^0$, and $\beta^0$ provided by a suitable initialization method (more details in Supplementary Section~D). In each iteration, the algorithm  descends along the gradient direction using step sizes \vspace{-1cm}

 \begin{equation}\label{eq:initial_step}
   \eta_{Z} = \frac{\eta}{2\|Z^0\|^2_\op}\,,\,\eta_{\alpha_S} = \frac{\eta}{4n}\,,\,\eta_{\alpha_{I -S}} = \frac{\eta}{4n_S}\,,\,\eta_{\beta} = \frac{\eta }{2\|S(X)\|^2_F}\,,  
 \end{equation}
for some positive constant $\eta$. 
At the end of each iteration, there is an additional projection step to ensure the estimates stay in suitably bounded sets for theoretical analysis. The specific choices of the constraint sets $\cC_Z, \cC_{\alpha}, \cC_{\beta}$ will be made clear later in the statements of the theoretical results. In practice, we simply perform $\mathcal{P}_{\cC_Z}(Z) = J Z$ and leave $\alpha$ and $\beta$ unchanged in step~\ref{eq:projection_alg}, which does not appear to affect empirical performance.


We remark here that the optimization problem  is closely related to the literature on binary (1-bit) matrix completion \citep{candes2010matrix, Koltchinskii2011, davenport20141}, where we observe a random subset of binary entries generated from a distribution with a low-rank structure. A key difference is that matrix completion typically assumes uniformly sampled entries, whereas in our setting edge visibility is governed by a random neighborhood $\cS$, inducing non-uniform and potentially biased observation; standard formulations also usually omit covariates. Nevertheless, both problems involve nonconvex optimization driven by low-rank structure. While some work uses nuclear norm relaxations \citep{davenport20141,candes2012exact}, these can be expensive due to repeated SVDs, whereas our projected gradient descent is computationally efficient and comes with theoretical guarantees we develop.

\section{Theoretical results}\label{sec:theoretical_results}

In this section, we establish bounds on the estimation errors and convergence rate for Algorithm~\ref{alg:lsp} and provide interpretations of these results in the context of partial information in a social network. We consider the model in~\eqref{eq:lsm} with parameters belonging to the space:
\vspace{-1.2cm}

\begin{multline}
  \cF_\Theta = \left\{ \Theta \mid \Theta = \alpha 1_n^\top  +1_n\alpha^\top  + \beta X + ZZ^\top , J^0Z = Z\right.\\
 \left. \max_{i\in [n]} \|Z_{i\cdot}\|^2 , 2\|\alpha\|_\infty, |\beta| \max_{i, j \in [n], i\neq j}|X_{ij}| \leq \frac{M_1}{3}, \max_{ i, j \in [n], i\neq j} \Theta_{ij} \leq -M_2 \right\}
\end{multline}
The bounds $M_1$, $M_2$, and the covariate $X$ can vary with the size of the network $n$. In what follows, we use stars to emphasize the true parameters, e.g., $\stTheta$, $\stG$, $\stZ$, $\stalpha$, and $\stbeta$. 

While it is natural to quantify the overall estimation error of $\stTheta$, we note that the missing submatrix $A_{I-S, I-S}$ makes estimating some parameters, for example, the corresponding component of the degree heterogeneity parameters $\stalpha_{I-S}$ more challenging, leading to a slower convergence rate. Consequently, the estimate for $\stTheta_{I-S,I-S}$ also converges a slower rate. Therefore, our analysis begins by focusing on the estimation error of $S(\stTheta)$. Notably, it is sufficient to upper bound this error by a weighted sum of the individual errors of $\stZ$, $\stalpha$, and $\stbeta$, defined as:
 $$e_t = \| \cstZ\|^2_\op\|\Delta_{Z^t}\|^2_\F + 2\|S(\Delta_{\alpha^t} 1^\top _{n})\|^2_\F + \|\Delta_{\beta^t} S(X)\|^2_\F\,,$$
given $Z^t, \alpha^t, \beta^t$ obtained from the $t$-th iterate. Further define $ \cstZ  = J \stZ$ 
using the centering matrix $J$ in Eq~\eqref{eq:J_form}  for the projection step in Algorithm~\ref{alg:lsp}. 
Let $\Delta_{Z^t} = Z^t - \cstZ R^t$, where $R^t = \arg \min_{R \in \cO(k)}\|Z^t - \cstZ R\|_\F$, and $\cO(k)$ is the set of $k \times k $ orthogonal matrices. Also denote $\Delta_{\alpha^t} = \alpha^t - \stalpha$, $\Delta_{\beta^t} = \beta^t - \stbeta$, $\Delta_{G^t} = Z^t (Z^t)^\top  - \stZ (\stZ)^\top  $, $\Delta_{\Theta^t} = \Theta^t - \stTheta$, representing deviations from the true parameters, and the partial error term $\Delta_{S(\Theta^t)} = S(\Theta^t - \stTheta)$. We will demonstrate that the error metric $e_t$ is closely related to $\|\Delta_{S(\Theta^t)}\|^2_\F$ and $\|\Delta_{\Theta^t}\|^2_\F$, and thus deriving the convergence guarantee for $e_t$ is the key to the analysis.

Below we begin by conditioning on the neighborhood set to separate partial‑information bias from generic estimation error and to identify neighborhood structures most favorable for estimation. Section~\ref{sec:lsp_performance} analyzes algorithm performance under ideal neighborhoods, while Section~\ref{sec:neighbor_model} treats general neighborhoods and quantifies the resulting bias. Finally, we extend the analysis to the full‑likelihood setting in Section~\ref{sec:random_neighbor_model}, in which the set $\cS$ is generated randomly according to the model.

\subsection{Theoretical results for a conditional likelihood with balanced neighborhood}\label{sec:lsp_performance}

We begin our analysis by conditioning on the neighborhood set $\cS$ of node 1, which is not only a natural consideration given the structure of the likelihood in Eq~\eqref{eq:likelihood}, but also allows us to relate and interpret the theoretical results with respect to the structure of $\cS$. In this case, $S$, $r_S$, and $B_{1j}$ for all $j \in [n]$, are given. The corresponding objective function (i.e., the negative log-likelihood function conditional on $\cS$) for optimization becomes
\begin{equation}\label{program:fixed} 
g(Z,\alpha,\beta) =  -\sum_{i \in \cS\backslash\{1\}, j > i}\left\{ B_{ij} \Theta_{ij} + \log\left(1 - \sigma(\Theta_{ij})\right)\right\}\,.
\end{equation}
We note here that the conditional version of the problem does not produce estimates for the parameters associated with node 1, such as $\alpha_{1}$ and $z_1$. 
For ease of exposition, we keep the current notations and set the estimation errors of $\stTheta_{1 \cdot}$ and $\stG_{1 \cdot}$ to 0.





Next we introduce some assumptions necessary for establishing the results. As the optimization problem is non-convex, the convergence of Algorithm~\ref{alg:lsp} requires a suitable initialization satisfying the following. 
\begin{assumption}[Initialization condition]\label{assp:Z_0} 
The initial values $Z^0, \alpha^0, \beta^0$ in Algorithm~\ref{alg:lsp} satisfy  $e_0 \leq c \gamma_S\|\cstZ\|^4_\op/e^{M_1}$  for some small enough constant $c$, where
\begin{equation}\label{eq:gamma_S}
    \gamma_S  = \min(r_S,(\kappa')^{-4} ) \qmq{and}\kappa' = \frac{\sigma_1(\cstZ)}{\sigma_k(\cstZ_S)}\,.  
\end{equation}
\end{assumption}
Recall that $\cstZ_S$ is a submatrix of $\cstZ$ for the neighbor nodes; $\gamma_S$ is closely related to the neighborhood ratio $r_S$, and more details will be explained later under more specific network models. An initialization algorithm based on universal singular value thresholding \citep{chatterjee2015matrix} is presented in Supplementary Section~D to meet the initialization requirement with theoretical justification.

The next assumption is related to the partial covariate matrix $S(X)$. It is easy to see that when $S(X) = 0$, $\beta$ cannot be estimated given only $B$.
When $S(X)$ is nonzero, the following assumption is necessary to ensure $\beta$ can be effectively estimated.

\begin{assumption}[Covariates] \label{assp:X} The stable rank of the partial covariate matrix $S(X)$ satisfies
$$r_\mathrm{stable}(S(X)) := \frac{\|S(X)\|^2_\F}{\|S(X)\|^2_\op} \geq C_1 \frac{k}{r^2_S}$$ for some large enough $C_1$. Also, we require  $r_\mathrm{stable}(X)  \geq C_2 k$ for some large enough $C_2$.
\end{assumption}

The condition on the partial covariate matrix $S(X)$ is analogous to that of the full covariate matrix $X$, as proposed in \citet{ma2020universal}. Assuming that the condition numbers of the matrices $S(X)$ and $ X$ are finite, Assumption~\ref{assp:X} holds when
$\mathrm{rank}(S(X)) \geq C_3 k/r_S^2$ and $\mathrm{rank}(X) \geq C_4 k$ for some large enough constants $C_3$ and $C_4$.  The lower bound on $\mathrm{rank}(X)$ is consistent with the requirement in \cite{ma2020universal} for estimating $\beta$ -- a larger rank is necessary as $k$ increases to ensure $X$ contains sufficient information for estimation. Under partial information, the lower bound on $\mathrm{rank}(S(X))$ becomes more stringent as $r_S$ approaches zero in sparse networks. 





As this section focuses on the conditional likelihood setting given $\cS$, we first consider a special type of neighborhood structure to simplify the analysis. 
\begin{assumption}[Strictly balanced neighborhood]\label{assp:strictly_balance} 
    The neighborhood $\cS$ of the centered individual satisfies $\sum_{j \in \cS} \stz_{j} =  \sum_{i \in [n]} \stz_{i} = 0$. 
\end{assumption}
Intuitively, an ideal $\cS$ should be representative of the overall network, so its neighborhood center should lie close to the network center at 0.
We refer to such a neighborhood as ``strictly balanced'', where the individual forms connections with others in the network in an unbiased way and the impact of their local view on estimation is less than in other settings. The assumption will be removed later.

 The subsequent lemma establishes 
 that $e_t$ upper bounds the estimations errors for parameters of interest, thus bounding the convergence rate for $e_t$ allows us to derive bounds on the respective parameter estimation errors. 

\begin{lemma}\label{lemma:et_theta_simple_v1}
  Conditional on the neighborhood $\cS$, under Assumptions~\ref{assp:X} and~\ref{assp:strictly_balance}, when $\|\Delta_{Z^t}\|_\F \leq c\|\stZ\|_\op$ for some constant $c>0$, we have
$
\|\Delta_{S(\Theta^t)}\|^2_\F \leq   C_1 (2+c)^{2}  e_t  $
for some constants $ C_1 > 0$.  Furthermore,
$
\|\Delta_{G^t}\|^2_\F \leq  (2+c)^2e_t \qmq{and}  \|\Delta_{\Theta^t}\|^2_\F 
 \leq C_2 \max \left\{( 2+ c)^2, r_S^{-1},   \frac{\|X\|^2_\F}{\|S(X)\|^2_\F}\right\} e_t     \,,$ for some $C_2 > 0$.
\end{lemma}

For a strictly balanced neighborhood, the following theorem establishes that with a suitable initialization, the error $e_t$ contracts at a rate comparable to a geometric series with an exponent
depending on the quantity $\gamma_S$, indicating at least linear convergence in the terminology of iterative methods.


\begin{theorem}[Conditional on strictly balanced neighborhood]\label{thm:nc_exact} 
We set the constraint sets in Algorithm~\ref{alg:lsp} as $\cC_Z= \{Z \in  \R^{n\times k} \mid J Z = Z\,,\, \max_{i\in [n]}\|Z_{i\cdot}\|^2 \leq M_1/3\}$, $
    \cC_\alpha = \{\alpha \in \R^n \mid 2\|\alpha\|_\infty \leq M_1/3\}$, and $\cC_\beta = \{ \beta \in \R \mid  |\beta| \max_{ i, j \in [n], i\neq j}|X_{ij}| \leq M_1/3\}$,
and set the step sizes as in Eq~\eqref{eq:initial_step} with $\eta < c$ where $c$ is a universal constant.
Conditional on the neighborhood $\cS$, under Assumptions~\ref{assp:Z_0},~\ref{assp:X} and~\ref{assp:strictly_balance}, 
suppose that  \vspace{-0.2cm}
\begin{equation}
 \|\stZ\|^2_\op \geq C_1 \gamma_S^{-1}  r_S^{-1/2} \sqrt{n} e^{M_1 - M_2/2}\max\{\sqrt{  e^{M_1} k}, 1\}   \vspace{-0.2cm}  \label{eq:Zop_cond}
 \end{equation}
for a sufficiently large constant $C_1$. Then there exist positive constants $\xi$, $C$ and $C'$  uniformly over $\cF_\Theta$, such that w.h.p.,
\begin{equation}\label{eq:nc_exact_prob_et}
       e_{t}  \leq  \left(1 - \frac{\eta\xi \gamma_S}{e^{M_1} }\right)^{t}e_0 +  \frac{C}{\xi \gamma_S r_S }\psi^2_n\,,\qmq{where} \psi^2_n = e^{2M_1}nk\max\{e^{-M_2},\frac{\log n}{n}\}\,,
    \end{equation}
    and for any $T \geq T_0 = \log\left(\frac{ e^{3M_1 - M_2}}{ \gamma_S^2 r_S M^2_1 n k}\right)/ \log\left(1 - \frac{\eta\xi \gamma_S}{e^{M_1} }\right)$,
\begin{align}
 & \|\Delta_{G^T}\|^2_\F\,,\, \|\Delta_{S(\Theta^T)}\|^2_\F  \leq C' \gamma_S^{-1} r_S^{-1}\psi^2_n\,,   \qmq{and} \label{eq:nc_exact_theta_SG}\\
   & \|\Delta_{\Theta^T}\|^2_\F  \leq C' \max\left\{ r_S^{-1},   \frac{\|X\|^2_\F}{\|S(X)\|^2_\F}\right\}\gamma_S^{-1} r_S^{-1}\psi^2_n \,. \label{eq:Theta_F} 
\end{align}
\end{theorem}



\begin{remark}
(i) As mentioned earlier, the missing submatrix $A_{I-S,I-S}$ implies the estimation of $\stTheta_{I-S,I-S}$ is more difficult. Consequently,  the overall estimation error $\|\Delta_{\Theta^T}\|^2_\F$ converges at a slower rate compared to the partial error $\|\Delta_{S(\Theta^T)}\|^2_\F$ by a factor of $\max\{r_S^{-1}, \frac{\|X\|^2_\F}{\|S(X)\|^2_\F}\}$.

(ii) Compared to Theorem 9 in \citet{ma2020universal}, the presence of $r_S$ and $\gamma_S$ in the results above reflects the effect of the neighborhood size and $\sigma_k(\stZ_S)$ on the convergence rate. In the special case that  $r_S$, $\kappa'$, $\frac{\|X\|^2_\F}{\|S(X)\|^2_\F}=\Omega(1)$ (i.e., the neighborhood is sufficiently large and captures representative information of the overall structure), we recover the same orders for the sparsity condition, contraction rate, and statistical precision $\psi_n^2$ for the parameters as in the full-network setting \citep{ma2020universal}.

\end{remark}

For the rest of this section, we use a specific network model as an illustrative example to provide deeper insight into the assumptions required for Theorem~\ref{thm:nc_exact} and the convergence rates. To this end, we consider the stochastic block model (SBM) \citep{holland1983stochastic}, which can be regarded as a submodel of the inner-product model.

Under the SBM, $n$ nodes are allocated to $K$ blocks or communities, and the probability of an edge between two nodes is associated with their community memberships. For illustration, let $K$ be fixed. $n_j$ represents the size of block $j$; $\cA_j$ represents the set of nodes in block $j$ for $j \in[K]$; the proportion of members in block $j$ is denoted by $r_j = n_j/n$. The SBM can be reparameterized into model~\eqref{eq:J0_form} as follows.

\begin{setting}[SBM]\label{set:DCSBM}
Under an SBM with $K$ blocks, the probability matrix $P$ has a block structure, which is preserved by the logit link function. Thus, we can write
$\stTheta = \mathrm{logit}(P) = U H U^\top$, where $U \in \R^{n\times K}$ is the community membership matrix -- for $i\in[n]$ and $j \in [K]$,  $U_{ij} = 1$ if $i \in \cA_j$, and 
$U_{ij} =  0$ if $ i \notin \cA_j$.  Here, $H \in \R^{K \times K}$ is a symmetric positive definite matrix with $\sigma_K (H
 ) \geq c$ and $\sigma_1 (H
 )/ \sigma_K (H
 ) \leq C$ for some constants $c,C>0$. Moreover, $\Theta$ can be written in the form of Eq~\eqref{eq:lsm}. Specifically, the corresponding latent position matrix $\stZ \in \R^{n\times k}$, with $k = K-1$, is obtained from the decomposition $J^0 U H U^\top J^0 = \stZ (\stZ)^\top$. Details of the reparameterization are provided in Supplementary Lemma~S3.
\end{setting}

We note that this setting is included only to illustrate the conditions of the general main theorems and error rates, rather than to propose new estimation methods for the SBM and its variants. Although well-established algorithms already exist for fitting the SBM (e.g., modularity based or spectral clustering \citep{karrer2011stochastic,J15}),  we develop a likelihood-based approach for a more general latent space model with covariates, for which existing SBM community detection methods do not directly apply.

Under the partial information framework, we denote the proportion of block $j$ members in the neighborhood as $r_{j,S} = | \cA_j \cap \cS|/n_S$ and the proportion of block $j$ members in the non-neighborhood set as $r_{j,I-S} = | \cA_j \backslash \cS|/n_{I-S} $. It is easy to see that Assumption~\ref{assp:strictly_balance} holds under the SBM when $r_j = r_{j,S}$. To simplify our calculations, we will further assume the community sizes are balanced such that 
\vspace{-0.23cm}
\begin{equation}\label{ex:simple_strictly}
   r_j, r_{j,S} = 1/K \qmq{for all} j \in [K]\,. \vspace{-0.2cm}
\end{equation}
The following lemma bounds the magnitude of $\gamma_S$ under the simple SBM setting above.
\begin{lemma}\label{lemma:set_1_kappa}
    Under Setting~\ref{set:DCSBM} and assuming Eq~\eqref{ex:simple_strictly} holds,   $\kappa' \leq c r_S^{-1/2}$ for some $c > 0$, implying that $\gamma_S$ is lower bounded by $r_S^2$. 
\end{lemma}
    We will show later that, in more general settings, $\kappa'$ is upper bounded by a term that depends on the expected neighborhood ratio with high probability. The lemma above and Eq~\eqref{ex:simple_strictly} allow us to simplify condition~\eqref{eq:Zop_cond} required in Theorem~\ref{thm:nc_exact} so that we can better interpret the results in terms of network sparsity and neighborhood ratio using the next corollary. 

\begin{Corollary}\label{corollary:kappa_even}
Consider the same assumptions, constraint sets $(\cC_Z$, $\cC_\alpha$, $\cC_\beta)$, and step sizes as in Theorem~\ref{thm:nc_exact}. Under Setting~\ref{set:DCSBM}  and assuming Eq~\eqref{ex:simple_strictly} holds, the results of Theorem~\ref{thm:nc_exact} (i.e., Eq~(\ref{eq:nc_exact_prob_et}-\ref{eq:Theta_F})) hold when 
\vspace{-0.2cm}
\begin{equation}\label{eq:DCSBM_balance_assump}
  r_S^{\frac{5}{2}} e^{(-3M_1 + M_2)/2} \geq C  n^{-\frac{1}{2}}\,, \vspace{-1em} 
 \end{equation}
which is a simplified condition derived from Eq~\eqref{eq:Zop_cond}.
\end{Corollary}

\begin{remark} \label{remark:sparsity_rs_discussion}
Using this corollary and Eq~\eqref{eq:DCSBM_balance_assump}, we can simplify the error bounds in Eq~(\ref{eq:nc_exact_prob_et}-\ref{eq:Theta_F}). 
For illustration, we further assume $\|X\|^2_\F / \|S(X)\|^2_\F = O(r_S^{-1})$ when investigating the bound for $\|\Delta_{\Theta^T}\|^2_\F$.
\begin{itemize}[leftmargin=1.2em, itemsep=0ex, topsep=0ex]
\item When the network sparsity is controlled with $e^{(-3M_1+M_2)/2} = \Omega(n^{-1/2})$ and the neighborhood size is large enough such that $r_S = \Omega(1)$, we have $n^{-1/2}$ convergence rate for the parameters with ${\|\Delta_{G^T}\|^2_\F}/{n^2} , {\|\Delta_{\Theta^T}\|^2_\F}/{n^2} = O_p(n^{-1/2})$.

\item For dense networks with $e^{-M_1} = \Omega(1)$, $r_S = \Omega(n^{-1/5})$ is sufficient to guarantee Eq~\eqref{eq:DCSBM_balance_assump}. 
Then, we have $ {\|\Delta_{G^T}\|^2_\F }/{n^2}= O_p(n^{- 4/7})$ and $ {\|\Delta_{\Theta^T}\|^2_\F}/{n^2} = O_p(n^{-3/7})$.
\end{itemize}


\end{remark}

\subsection{Theoretical results for a conditional likelihood with general neighborhood} \label{sec:neighbor_model}

Assumption~\ref{assp:strictly_balance} serves as a convenient point to start our analysis but is often not satisfied in practice. Therefore, in what follows, we consider the more general case without this assumption, i.e., under $J\stZ \neq \stZ$. 
We assume that the individual does not have a full network view, with $r_S < c$ for some positive constant $c < 1$. To quantify the influence of neighborhood structure, we introduce the following measure to capture the amount of ``imbalance" within $\cS$,
\begin{equation}\label{eq:US_form}
  U^2_{S} = \frac{1}{n}\sum_{i \in [n]} ( {\stz_i}^\top\sum_{j \in \cS}   \stz_j )^2\, . 
\end{equation}
When Assumption~\ref{assp:strictly_balance} holds so that the center of $\cS$ aligns with that of the entire network (i.e., $\sum_{j \in \cS} \stz_j = 0$), $U_{S}$ reduces to zero. The more the center of the neighborhood deviates  from the origin, the larger $U_{S}$ becomes, indicating a greater amount of imbalance. We further normalize $U_{S}$ by $\|\stG\|_\F$ and use $U_{S}/\|\stG\|_\F$ as a measure of neighborhood imbalance. As we will demonstrate later in theoretical and numerical analysis, this measure is closely related to the estimation errors and reflects the quality of the available information in the partial view of $\cS$. 


The rest of this section provides convergence results for Algorithm~\ref{alg:lsp} conditional on a general neighborhood $\cS$, removing Assumption~\ref{assp:strictly_balance}. Similar to Lemma~\ref{lemma:et_theta_simple_v1}, the following lemma establishes upper bounds for various estimation errors in terms of $e_t$.

\begin{lemma}\label{lemma:et_theta_all_compare_imbalance}
Conditional on $\cS$, under Assumption~\ref{assp:X}, when $\|\Delta_{Z^t}\|_\F \leq c\|\cstZ\|_\op$, we have
$
\|\Delta_{G^t}\|^2_\F,  \|\Delta_{S(\Theta^t)}\|^2_\F \leq (1 + c_0)(2+ c)^2 e_t  +  \frac{C_0}{r_S} U^2_{S}$ and
$
 \|\Delta_{\Theta^t}\|^2_\F 
 \leq (1 + c_0) \max \left\{(2+ c)^2, r_S^{-1},   \frac{\|X\|^2_\F}{\|S(X)\|^2_\F}\right\} e_t + \frac{C_0}{r_S} U^2_{S}$,
 for some $0<c_0<1$ and $C_0 > 0$.
\end{lemma}
The main difference from Lemma~\ref{lemma:et_theta_simple_v1} is the term involving $U_S$, which can be viewed as a form of estimation bias arising from a general neighborhood $\cS$, as we demonstrate in the next theorem. 

\begin{theorem}[Conditional on general neighborhood]\label{thm:nc_exact_imbalanced} 
Consider the same constraint sets $(\cC_Z$, $\cC_\alpha$, $\cC_\beta)$ and step sizes as in Theorem~\ref{thm:nc_exact}. Conditional on $\cS$, under
Assumptions~\ref{assp:Z_0}-~\ref{assp:X}, 
suppose 
\begin{equation}\label{eq:thm:nc_exact_imbalanced_require}
\|\cstZ\|^2_\op \geq C_1 \gamma_S^{-1} r_S^{-1/2} e^{M_1 - M_2/2}  \sqrt{n} \max\left\{\sqrt{e^{M_1} k}, \sqrt{U^2_{S}/( n e^{-M_2})}\right\} 
\end{equation} 
for a sufficiently large constant $C_1$. Then there exist positive constants $\xi$, $C$ and $C'$ uniformly over $\cF_\Theta$, such that w.h.p.,
    \begin{equation} \label{eq:et_imbalance_deter} 
       e_{t}  \leq \left(1 - \frac{\eta\xi \gamma_S}{e^{M_1}}\right)^{t}e_0  + \frac{C}{\xi \gamma_S r_S}\left(  \psi_n^2 + e^{M_1}U^2_{S}\right)\,,
    \end{equation}
and for any $T \geq T_0 =  \log\left(\frac{ e^{3M_1 - M_2}}{ \gamma_S^2 r_S M^2_1 n k}\right)/ \log\left(1 - \frac{\eta\xi \gamma_S}{e^{M_1} }\right)$,
\begin{equation*}
\|\Delta_{G^T}\|^2_\F,  \|\Delta_{S(\Theta^T)}\|^2_\F\leq  C'\gamma_S^{-1} r_S^{-1}\left(  \psi^2_n +  e^{M_1}{U^2_{S}}\right) \qmq{and}
 \end{equation*}
 \vspace{-1cm}
 \begin{equation}\label{eq:us_effect}
 \|\Delta_{\Theta^T}\|^2_\F  \leq  C'\max\left\{ r_S^{-1},   {\|X\|^2_\F}/{\|S(X)\|^2_\F}\right\}\gamma_S^{-1} r_S^{-1}\left(  \psi^2_n +  e^{M_1}{U^2_{S}}\right)\,.
\end{equation}  
\end{theorem}

We note that when $U_S = 0$, the theorem above recovers the results in Theorem~\ref{thm:nc_exact}. Thus, we refer to the extra error term involving $U^2_{S}$ as the bias arising from the individual's local view of the full network.

\begin{remark}
The magnitude of the bias term can overwhelm the statistical precision term $\psi_n^2$, depending on the structure of $\cS$. Consider a block model as in Setting~\ref{set:DCSBM} with two communities, $\stZ \in \R^{n\times 1}$, such that $\stz_i = 1$ for $i\in\cA_1$, $|\cA_1| = n_1$, and $\stz_i = -n_1/(n-n_1)$ for $i\in\cA_2$. If $\cS = \cA_1$, that is the individual's view is extremely polarized with friends only from one community, it is easy to check that $U^2_{S}   \geq   r_S^2 n^2$. In this case when $ e^{- M_1 + M_2} r_S^2  \gg k n^{-1} $, we have $e^{M_1}U^2_S \gg \psi^2_n$ in the error bounds of Eq~\eqref{eq:et_imbalance_deter}--\eqref{eq:us_effect}. 


\end{remark}

\begin{figure}
    \centering
      \begin{subfigure}[b]{0.3\textwidth}
          \includegraphics[width = \textwidth]{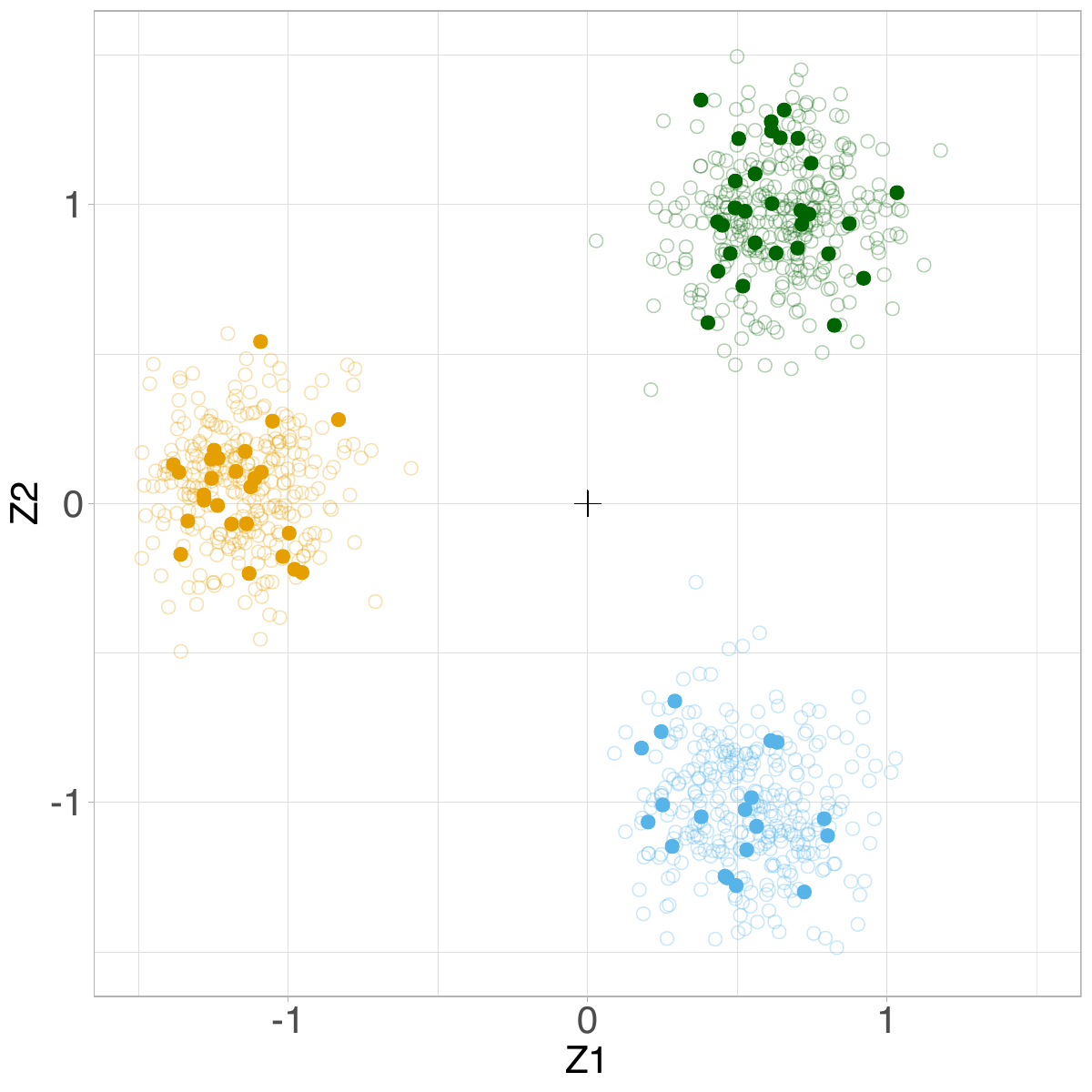}
    \caption{ $U_S/\|\stG\|_\F = 0.009$} \label{fig:ill_balance}
    \end{subfigure}
    \hspace{2cm}
    \begin{subfigure}[b]{0.3\textwidth}
          \includegraphics[width = \textwidth]{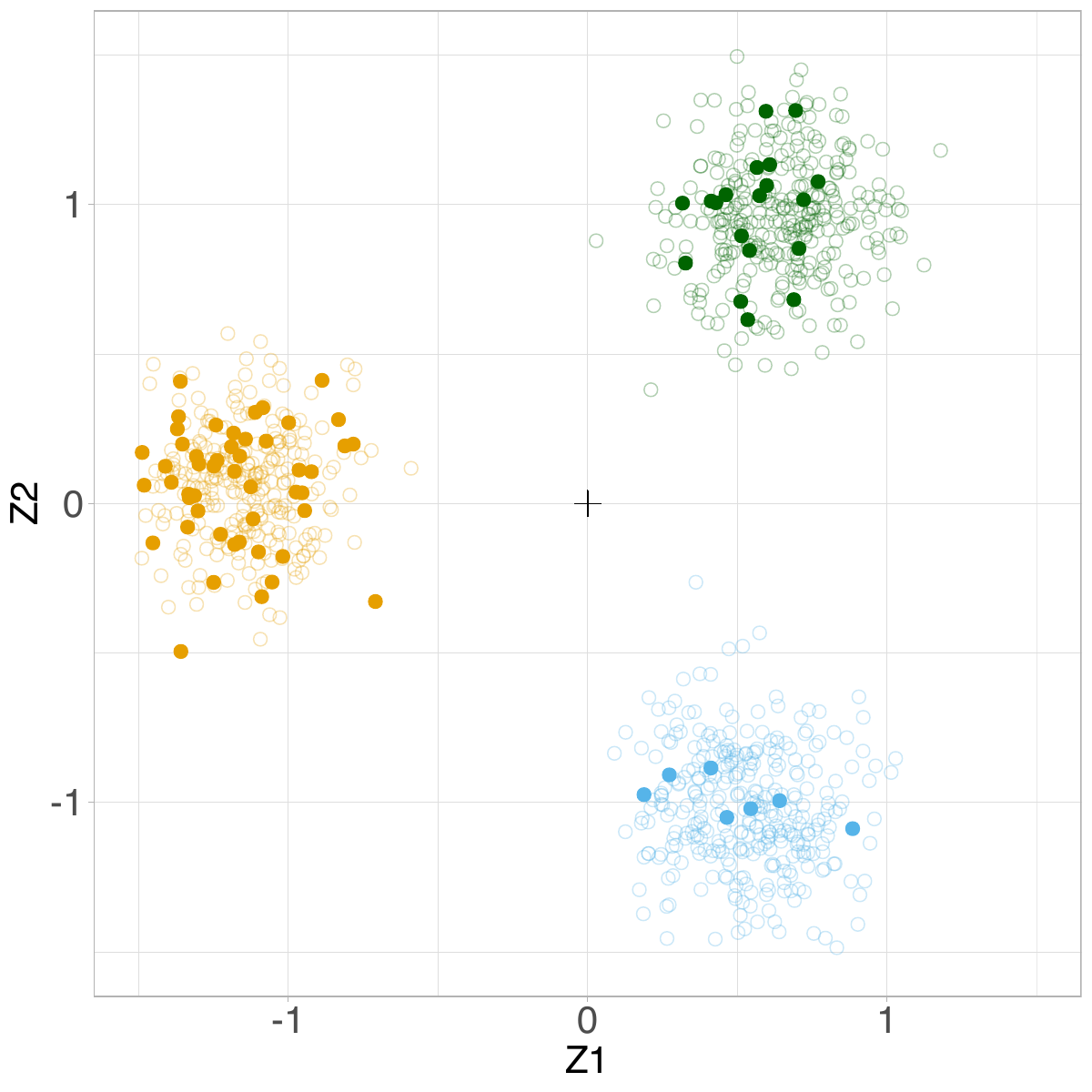}
    \caption{$U_S/\|\stG\|_\F = 0.044$} \label{fig:ill_imbalance}
    \end{subfigure}
    \caption{Different types of neighborhood: (a) almost balanced; (b) imbalanced, and their corresponding imbalance measures.  Solid circles represent direct neighbors of the centered node, while empty circles represent the other nodes in the network. 
    }\label{figure:ill_neighbor_structure}
\end{figure}

Finally, we illustrate neighborhood imbalance with a simulated example. Figure~\ref{figure:ill_neighbor_structure} shows two neighborhoods around a node in a network with 
$n=1000$ and latent positions drawn from a three-component mixture. In Figure~\ref{fig:ill_balance}, edges are formed with uniform probability across nodes, producing an almost balanced neighborhood with $U_S$ close $0$.
In Figure~\ref{fig:ill_imbalance}, neighbors are determined by the realized adjacency matrix; the node connects mostly within its own cluster, yielding a larger imbalance measure. Simulation details are given in Supplementary Section~A.1.


\spacingset{1}
\spacingset{1.9}

\subsection{Theoretical results for the full likelihood}\label{sec:random_neighbor_model}

Now we broaden our scope to consider the full likelihood function and allow $\cS$ to be random. In this general case, rather than interpreting the results for a specific neighborhood, we consider the average connection probability surrounding the centered node. More specifically, recall that the edges surrounding node 1 are distributed as $A_{1i} \sim \mbox{Bernoulli}(P_{1i})$ and $P_{1i} = \sigma(\Theta_{1i})$ for $i \neq 1$.  We define the average connection probability around node 1 as $p_S = \frac{1}{n} \sum_{i \in [n]}P_{1i}$, which can also be interpreted as the expected neighborhood ratio. We assume that $p_S < c$ for some positive constant $c < 1$.
As $r_S$, $\cS$, $J$ (consequently $\cstZ$) are all random, we need the following assumptions analogous to Assumptions~\ref{assp:Z_0} and~\ref{assp:X}. 

\begin{assumption} [Analogous to Assumption \ref{assp:Z_0}]\label{assp:Z_0_p}
The initial values $Z^0, \alpha^0, \beta^0$ in Algorithm~\ref{alg:lsp} satisfy w.h.p. $e_0 \leq c p_S^2n^2/(e^{M_1} k^4)$  for some small enough constant $c$.
\end{assumption}

\begin{assumption}[Analogous to Assumption \ref{assp:X}]\label{assp:X_p} 
We assume that $r_{\mathrm{stable}}(S(X)) \ge C_1 k/p_S^2$ w.h.p. for sufficiently large $C_1$, and $r_{\mathrm{stable}}(X) \ge C_2 k$ for sufficiently large $C_2$.
\end{assumption}

The following lemma provides high probability bounds on the neighborhood ratio $r_S$ and neighborhood imbalance $U^2_{S}$ as defined in Eq~\eqref{eq:US_form}.

\begin{lemma}\label{lemma:prob_U} We assume that $p_S \geq n^{-c_1}$ for some constant $c_1 \in (0,1)$. Then, uniformly over $\cF_\Theta$,  w.h.p., $ r_S = \Omega(p_S)  $ and 
\begin{equation}\label{eq:prob_U_general}
  U^2_{S} =  \frac{1}{n}\sum_{i \in [n]}  ({\stz_i}^\top\sum_{j \in \cS}   \stz_j )^2 = O\left( \frac{\|\stZ\|^2_\op}{n}  \max\left\{   k n p_S  M_1 \log(n), \delta^2_{n}\right\}\right), \mbox{ where }  \delta^2_{n} = \| \sum_{i \in [n]}P_{1i} \stz_{i} \|^2 \,.
\end{equation}
\end{lemma}
\vspace{-0.5cm}




\begin{remark}
When the centered individual has uniform edge probabilities with all nodes in the network,  i.e., $P_{1i} = p_S$ for all $i \in [n]$, we have $\delta_n^2 = 0$, leading to a simplified upper bound. More specifically, when an individual is equally likely to form connections with everyone else in the network, the upper bound for $U_S$ depends on the neighborhood $\cS$ solely through the sparsity level, as determined by $p_S$, and is independent of the latent positions of the neighbors.


\end{remark}

We need an additional assumption on $\kappa'$ and $\cstZ$ under the full likelihood setting. 

\begin{assumption}[Latent position condition]\label{assp:kappa_p_bound}
We assume that $\kappa' = O(\sqrt{k/p_S})$ and $\|\cstZ\|^2_\op = \Omega(n/k)$ w.h.p. 
\end{assumption}

Assumption~\ref{assp:kappa_p_bound} is mild and holds under common network models, as demonstrated below.

\begin{example}\label{eg:kappa_p_bound}
Consider a SBM in Setting~\ref{set:DCSBM}. We denote $p_{S,j} = \frac{1}{n_j} \sum_{i \in \cA_j} P_{1i} $ for all $j \in [K]$, which is the average probability of edges with the $j$-th community. It follows then $p_S =\sum_{j \in [K]} r_j p_{S,j}$. We assume that the proportion of the $j$-th community, $r_j$, is lower bounded by a constant for all $j \in [K]$ and $p_{S,j}$ is of the same order as $p_S$ with $p_{S,j} \simeq p_S$. Supplementary Lemma~S13 shows that Assumption~\ref{assp:kappa_p_bound} holds under this example.
\end{example}

Lemma~\ref{lemma:prob_U} and Assumption~\ref{assp:kappa_p_bound} immediately imply that w.h.p.  $\gamma_S = \Omega(p_S^2/k^2)$. Using these bounds, we are now ready to establish the high-probability upper bound for each iteration of Algorithm~\ref{alg:lsp} under the full likelihood model. 

\begin{theorem}[Full likelihood]\label{thm:nc_exact_random} 
Consider the same constraint sets $\cC_Z$, $\cC_\alpha$, $\cC_\beta$ and step sizes as in Theorem~\ref{thm:nc_exact}. Define 
$\cU^2_{S} =  \frac{\|\stZ\|^2_\op}{n}  \max\left\{   k n p_S  M_1 \log(n), \delta^2_{n}\right\}$.
Under Assumptions~\ref{assp:Z_0_p},~\ref{assp:X_p} and~\ref{assp:kappa_p_bound},  suppose that \vspace{-0.2cm}
\begin{equation}\label{eq:cU_S_require}
p^{5/2}_S e^{(-3M_1 + M_2)/2}\min\{\sqrt{ne^{M_1-M_2} k} /\cU_{S},1\}\geq C_1 k^{7/2} n^{-\frac{1}{2}} \vspace{-0.2cm}
\end{equation}
for a sufficiently large constant $C_1$. Then there exist positive constants $\xi$, $C$, $C'$ uniformly over $\cF_\Theta$, such that w.h.p.,
\vspace{-0.3cm}
     \begin{equation}\label{eq:thm_3_1}
       e_{t}  \leq  \left(1 - \frac{\eta\xi p_S^2 }{e^{M_1} k^2}\right)^{t}e_0 +  \frac{C k^2 }{\xi p_S^3}\left(\psi^2_n + e^{M_1} \cU^2_{S}\right)\,,\vspace{-0.2cm}
    \end{equation}
and for any $T \geq T_0 = \log\left(\frac{ e^{3M_1 - M_2}k^7}{p_S^5  n }\right)/ \log\left(1 - \frac{\eta\xi p_S^2 }{e^{M_1} k^2}\right)$, \vspace{-0.2cm}
\begin{equation}\label{eq:thm_3_2}
 \|\Delta_{G^T}\|^2_\F\,,\, \|\Delta_{S(\Theta^T)}\|^2_\F \leq C'\frac{k^2 }{p_S^3}\left(\psi^2_n + e^{M_1} \cU^2_{S}\right)\,. \vspace{-0.2cm}
\end{equation}
By further assuming   w.h.p.  $\|X\|^2_\F/\|S(X)\|^2_\F = O(p_S^{-1})$, we have  w.h.p., \vspace{-0.2cm}
\begin{equation}\label{eq:thm_3_3}
\|\Delta_{\Theta^T}\|^2_\F  \leq C'\frac{k^2 }{p_S^4}\left(\psi^2_n + e^{M_1} \cU^2_{S}\right)\,.
\end{equation}
\end{theorem}

\vspace{-0.5cm}


\begin{remark} \label{remark:error_rate_small_big_delta}


To examine the effect of $\cU^2_S$, assume that the local density $p_S \leq n^{-c_2}$ for some $c_2 \in (0, 1)$ for convenience.

\begin{itemize}[leftmargin=1.2em, itemsep=0ex, topsep=0ex]
    \item  When $\delta_n = O(\sqrt{e^{M_1-M_2}} / \|\stZ\|_\op)$, the condition Eq~\eqref{eq:cU_S_require} essentially reduces to the high-probability version of 
Eq~\eqref{eq:DCSBM_balance_assump}, 
which is required under the strictly balanced setting. The estimation errors are bounded by $\|\Delta_{G^T}\|^2_\F = O(\psi_n^2/p_S^3)$ and $\|\Delta_{\Theta^T}\|^2_\F = O(\psi_n^2/p_S^4)$ with high probability, which are consistent with Eq~\eqref{eq:nc_exact_theta_SG}--\eqref{eq:Theta_F} of Theorem~\ref{thm:nc_exact}. In other words, when $\delta_n$ is small enough, the individual bias $\cU_S$ is small and has a negligible effect on the algorithm's convergence behavior and estimation accuracy.

   \item When $\delta_n = \Omega(\sqrt{e^{M_1-M_2}} / \|\stZ\|_\op)$, the term $e^{M_1} \cU^2_{S}$, which arises from individual bias, will be greater than the original statistical precision $\psi^2_n$ and dominate the estimation errors in Eq~\eqref{eq:thm_3_1}--\eqref{eq:thm_3_3}.
\end{itemize}

\begin{remark}
     If we assume $e^{-M_1} \asymp e^{-M_2} \asymp p_S \asymp \rho_n$ for a sparsity parameter $\rho_n$ and $\delta_n = O(\sqrt{e^{M_1-M_2}} / \|\stZ\|_\op)$, the sparsity requirement in Eq~\eqref{eq:cU_S_require}  reduces to \ $\rho_n \geq n^{-1/7}$. This condition is more stringent than  that of the full network setting in \cite{ma2020universal} (Theorem 9, which requires $\rho_n \geq n^{-1/3}$). A similar phenomenon appears under block models: \cite{han2020individual} requires $\rho_n \geq n^{-1/2}$ for community detection from partial information, whereas typical full-network results can hold under much weaker sparsity, e.g., $\rho_n \gtrsim (\log n)/n$ \citep{abbe2015exact}. Thus, the partial information setting imposes a roughly square-root more stringent sparsity requirement. 
\end{remark}

\end{remark}

\vspace{-2em}

\section{Simulation studies}\label{sec:simulation_study}

In this section, we present simulation results to demonstrate the performance of our algorithm in estimation in various settings. All the results here are based on $100$ repetitions.

 \vspace{-0.2cm}

\begin{simulation}\label{sim:setting_general}
The parameters in Model~\eqref{eq:lsm} are generated as follows:
\begin{itemize}[leftmargin=1.2em, itemsep=0ex, topsep=0ex]
    \item Degree heterogeneity parameters: we set $\stalpha_i = - c_a  n 
 a_i/( \sum_{j \in [n]} a_j)$ for $i \in [n]$, where $a_j \overset{iid}{\sim} \mathrm{Uniform}(1,3)$ and  $c_a = 1$.
 
\item Latent positions: we generate $\mu_i \in \R^k$ with entries i.i.d. $ \mathrm{Uniform}(-1,1)$, and sample rows of $U \in \R^{n\times k}$ as $U_{i\cdot} \overset{iid}{\sim} \mathrm{Normal}(\mu_i, I_k)$, $i \in [n]$.
We then compute $\stG = n G/\|G\|_\F$ with $G = J^0 U U^\top J^0$. 

\item Covariate part: we set $\stbeta =- 0.5$  and generate $V\in \R^{n \times n}$ with entries $V_{ij} =  \min(|v_{ij}|, 2)$, where $v_{ij} \overset{iid}{\sim} \mathrm{Normal}(1,1)$, $ i,j \in [n]$. The covariate matrix is computed as $X = n V/ \|V\|_\F $.

\end{itemize}

\end{simulation}

\begin{figure}[!ht]
\centering
\begin{subfigure}[b]{0.33\textwidth}
\centering
\includegraphics[width=\textwidth]{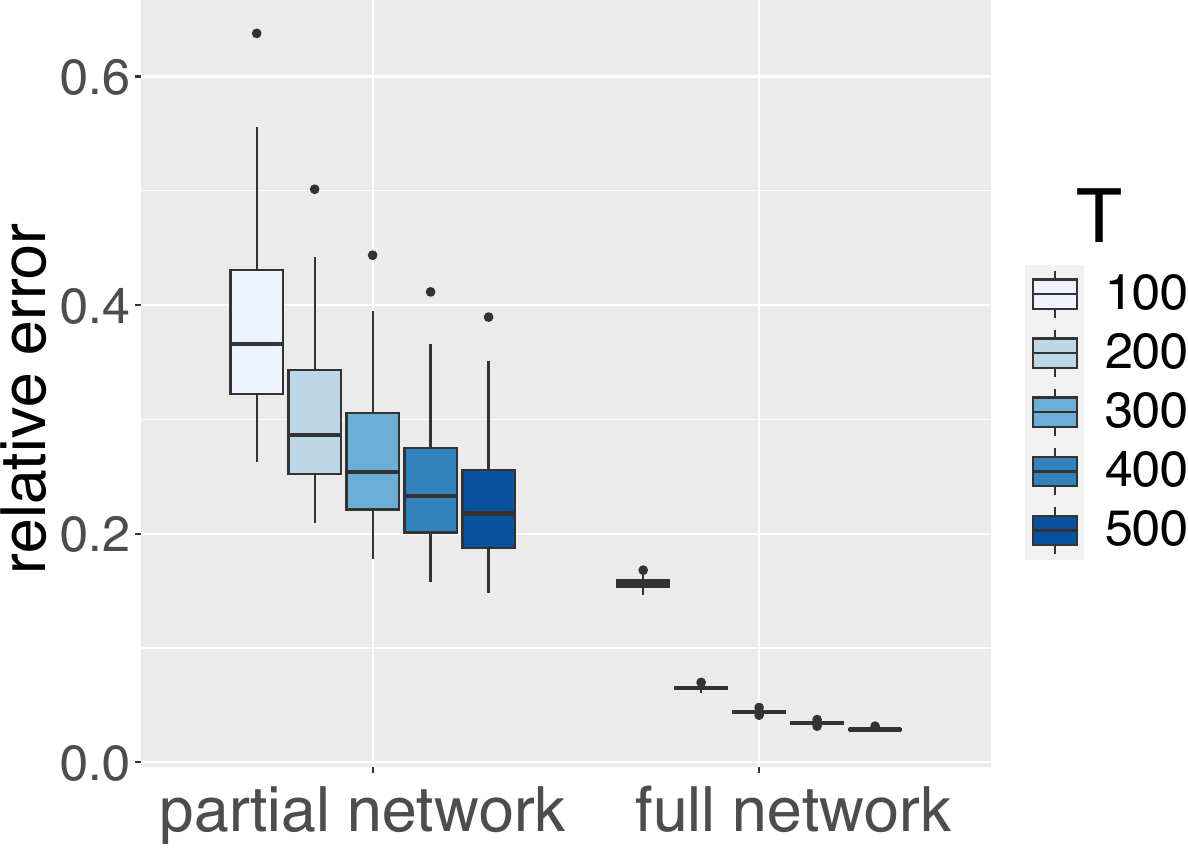}
\caption{}\label{fig:convergence_rate_general}
\end{subfigure}
\hspace{4em}
    \begin{subfigure}[b]{0.33\textwidth}
    \centering
\includegraphics[width=\textwidth]{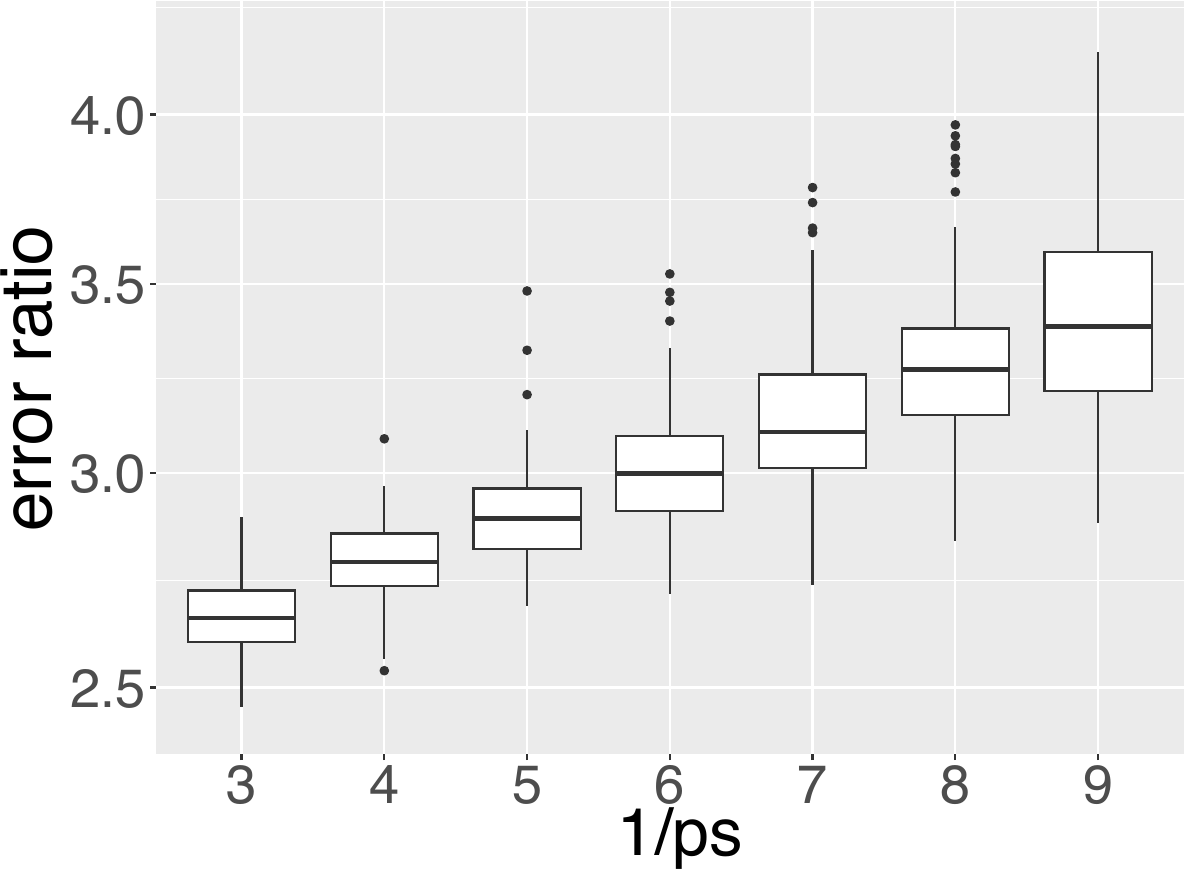}
    \caption{ }\label{fig:Theta_change}
 \end{subfigure}
 \vspace{-0.3cm}
 \caption{  (a) Convergence of the relative error of $\Theta$ for partial information and the full network as $T$ increases in Simulation~\ref{sim:setting_general}; (b) Change in the distribution of the error ratio $\|\Delta_{\Theta^\mathrm{part}}\|^2_\F/\|\Delta_{\Theta^\mathrm{full}}\|^2_\F$ as $p_S$ varies in Simulation~\ref{sim:setting_general_1} (y‑axis uses a $y^{1/4}$ scale)} \label{fig:G_Theta_change}
\end{figure}

We set $n = 1000$, $k = 4$ and generate independent copies of the adjacency matrix $A$ using the above setting.  In Algorithm~\ref{alg:lsp}, we set $\eta = 0.2$, $T = \{100, 200, \ldots, 500\}$ and use Algorithm~S1 to initialize values of the parameters. The estimation and convergence results as $T$ increases are presented in Figure~\ref{fig:convergence_rate_general}. We measure the performance of our algorithm by computing $\|\Delta_{\Theta^T}\|^2_\F/\|\stTheta\|^2_\F$ as the relative error of estimating $\Theta$ (using the partial network centered at node 1) and compare it with that of using the full network. As expected, the full-network estimates result in smaller errors.

We further compare the partial‑information and full‑network estimates for different values of neighborhood ratio $p_S$, using the following setting.

 \vspace{-0.2cm}

\begin{simulation}\label{sim:setting_general_1}
This setting retains all
the parameter settings of Simulation~\ref{sim:setting_general},  except for $\stalpha$. Here $\stalpha$ takes the same form as in Simulation~\ref{sim:setting_general}, except that the constant $c_a$ was chosen via binary search so that the average connection probability, $\frac{1}{n^2}\sum_{i,j \in [n]} P_{ij}$, is approximately the desired $\rho_n$ level. Moreover, to vary the  neighborhood connection probability $p_S$, we choose $\stalpha_1$ via binary search so that
$\frac{1}{n}\sum_{j \in [n]} P_{1j} = p_S$. By keeping the other parts of the model fixed, we can directly examine the relationship between the estimation errors and $p_S$ (or $\rho_n$). 
\end{simulation}

We examine the effect of $p_S$ by fixing $\rho_n = 0.2$, $n = 500$, $k = 4$, $T = 100$, and $\eta = 0.2$ and varying $p_S = 1/c_s$ with $c_s \in \{3,4,\ldots,9\}$.  We compute the estimation error ratio $\|\Delta_{\Theta^\mathrm{part}}\|^2_\F/\|\Delta_{\Theta^\mathrm{full}}\|^2_\F $, where $ \Delta_{\Theta^\mathrm{full}}$ and $\Delta_{\Theta^\mathrm{part}}$ denote the estimation errors from the full-network and partial-information settings, respectively. As shown in Figure~\ref{fig:Theta_change}, the error ratio increases approximately linearly with $1/p_S^4$, broadly in line with the rate  in the first scenario of Remark~\ref{remark:error_rate_small_big_delta}. Supplementary Figure~S2 indicates a low imbalance level in this simulation, suggesting negligible local-view bias and supporting this regime. A similar plot for the error ratio of $G$ is shown in Supplementary Figure~S1. 


Additional simulation results, examining differential convergence rates for different parameter components and evaluating the effectiveness of the proposed step sizes and centering matrix, are provided in Supplementary Section~A.2.  Furthermore, another simulation setting  generate networks with the latent positions following a two-component mixture is included in Supplementary Section~A.2 with estimation and convergence results in Supplementary Figure~S8.

\spacingset{1}

\spacingset{1.9}

\section{Real data analysis}\label{sec:real_data}

\subsection{ Zachary’s karate club data}\label{sec:real_data_karate}

\begin{figure}
 \hspace{-1cm}
 \begin{minipage}{0.3\linewidth}
		\includegraphics[width=4cm]{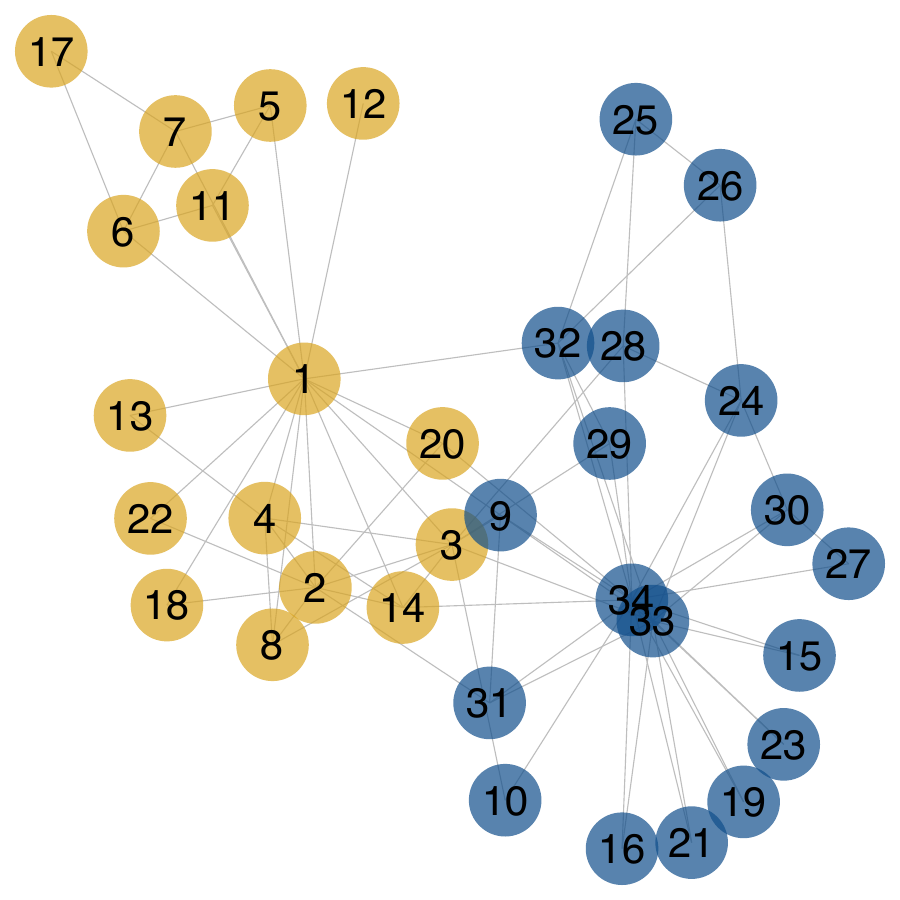}
	\end{minipage}
	\begin{minipage}{0.6\linewidth}
		\centering
			\footnotesize
		\begin{tabular}{c|cccccc}
  \hline\hline
 Node & 1 & 2 & 3 & 20  &  32 & 34\\\hline
              Degree & 16 & 9 & 10 & 3 & 6 & 17\\
            Fraction of observed edges& 0.654 & 0.513 & 0.705 & 0.526 & 0.654 & 0.641\\
              Betweenness & 0.438 & 0.054 & 0.144 & 0.032 & 0.138 & 0.304 \\
            Eigenvector & 0.355 & 0.266 & 0.317 & 0.148 & 0.191 & 0.373\\
            Closeness &  0.569 & 0.485 & 0.559 & 0.500 & 0.541 & 0.550\\    
                Imbalance &0.344 & 0.232 & 0.144 & 0.055 & 0.072 & 0.353\\
            \hline 
            Accuracy (ours) &0.618 & 0.794 & 0.971 & 0.882 & 0.853 & 0.588\\ 
              
            Accuracy \citep{han2020individual}&    0.529 &   0.706 &    0.941 &    0.941 &   0.824 &    0.738\\ \hline 
		\end{tabular}
	\end{minipage}
 \caption{ Left panel: the karate club network with nodes color-coded according to ground truth community labels. Right panel: clustering accuracy using different nodes as the center (second‑to‑last row for our method and last row for \cite{han2020individual}), and node-specific attributes, including node degree, fraction of edges observed in their partial network, commonly used centrality measures,  and the imbalance measure. 
}\label{fig:karate}
 \end{figure}

The karate club network \citep{zachary1977information} (Figure~\ref{fig:karate}, left panel) consists of 34 nodes and two communities that formed after conflicts between the instructor (node 1) and the administrator (node 34). While this network is typically analyzed from a global perspective using the full network, we study it under partial information by applying Algorithm~\ref{alg:lsp} with $k=2$ and different nodes as the center. We first estimate the latent positions of all nodes in the network, followed by community detection using k-means clustering. Although our primary focus is latent position estimation, clustering accuracy provides a useful secondary metric for evaluation when ground-truth latent positions are unavailable, as is standard in the literature (e.g., \citep{krivitsky2008fitting, ma2020universal, gao2022community}).

\spacingset{1}
 \spacingset{1.9}


The right panel of Figure~\ref{fig:karate} compares the clustering accuracy for different nodes with node-specific attributes such as degree, fraction of edges observed, and other commonly used centrality measures including betweenness, closeness \citep{freeman2002centrality}, and eigenvector centrality \citep{bonacich1987power}. When computing the imbalance measure, we use the entire network to obtain estimates of the latent positions and calculate empirical estimates of the relative imbalance measure $U_S/\|\stG\|_\F$. All these attributes measure the importance of a node based on its connectivity patterns. We note that generally nodes with higher centrality measures tend to be more influential in the network, hence are likely to capture more information from their neighborhood for estimation. On the other hand, a smaller imbalance measure is more favorable for improved estimation, as explained in the previous sections. In particular, as shown in the table of Figure~\ref{fig:karate}, nodes 20 and 32, despite their small degrees and low fractions of observed edges, have low imbalance measures and outperform the two hub nodes (nodes 1 and 34) in community detection accuracy. We compare with \citet{han2020individual}, which performs community detection directly on the $L=2$ partial information network. The performance pattern is similar across nodes, and our method is more accurate in four of six cases.

Supplementary Figure~S10 shows that the imbalance measure is most strongly associated with clustering accuracy among commonly used centrality measures and is also largely uncorrelated with them, indicating it captures complementary information.

\subsection{
U.S. Congress co-sponsorship networks}\label{subsec:congress}

We consider the legislative cosponsorship data in the U.S. House, which records the sponsor and co-sponsor of each bill. The dataset collected and discussed in \citet{fowler2006legislative} spans the period from 1973 to 2004. We divide the dataset into several five-year periods and construct a co-sponsorship network for each period; in each network, nodes represent legislators, and an edge exists between two legislators if their co-sponsorship instances surpass the median co-sponsorship frequency for all legislator pairings of that period. Legislators with no connections are removed from that period.

We apply Algorithm~\ref{alg:lsp} with $k=2$ to estimate latent positions in the whole network, using different legislators as the center. Taking the period 1990-1994 as an example, the entire network has 544 nodes with their estimated positions visualized in Figure~\ref{fig:85_90_congress_1} . We find that quite often, using nodes with moderate degrees leads to estimates that better align with Figure~\ref{fig:85_90_congress_1}. Nodes with higher degrees tend to be associated with higher imbalance measures (Supplementary Figure~S13), leading to potentially biased estimates. On the other hand, low-degree nodes may fail to reach a significant proportion of the nodes in the network within their partial view, resulting in insufficient information for estimation. In Figures~\ref{fig:85_90_congress_2}--\ref{fig:85_90_congress_3}, Armey has a degree of $208$ and imbalance measure of $0.185$, while Furse has a degree of $234$ and imbalance measure of $0.133$. The estimates from Furse's partial network perform better, as they more closely resemble the positions estimated using full network information.

More visualizations of the results for additional legislators and their local views can be found in Supplementary Figure~S12. These legislators also have degrees around $200$ but with varying imbalance measures, as shown in Table~\ref{table:inf_congress_example}. As before, local views with lower imbalance measures tend to yield better estimation results, as quantified by the evaluation metrics in the last two columns of Table~\ref{table:inf_congress_example}. The second‑to‑last column reports the relative error between the latent position estimate $\widehat G^\mathrm{part}$ obtained from partial information and the full-network estimate $\widehat G^\mathrm{full}$, computed as $\|\widehat G^\mathrm{part} - \widehat G^\mathrm{full} \|^2_\F/ \| \widehat G^\mathrm{full} \|^2_\F$. The last column reports in‑sample AUC values under model~\eqref{eq:lsm} using partial‑information estimates of $\Theta$. The full‑network estimator attains an AUC of $0.995$. All local‑view estimates achieve reasonably good model fit, with AUC values above $0.9$.

A closer look at these legislators with more balanced local view  suggests that they began their terms relatively recently around the observed period (1990–1994) and tend to have more balanced connections across both parties Table~\ref{table:inf_congress_example}. This finding suggests that studies on social connections in the U.S. Congress could benefit more from studying new members who have moderate degrees and relatively low imbalance measures, rather than solely focusing on well-established legislators with high degrees. The results for other periods and legislators, in particular, the three legislators with high degrees and used as reference points in the visualizations (Crane, Young, and Pelosi) are provided in Supplementary Figures~S14-S17.

\spacingset{1}

\begin{table}[!ht]
\footnotesize
    \centering
\begin{tabular}{c|cccc|cc}
    \hline \hline
Legislator & Degree & Imbalance &Political party & In office & Relative error &  AUC\\ \hline
Richard Armey & 208 & 0.185  & Republican & 1985-2003 & 0.798 &  {  0.910}\\
Elizabeth Furse &  234 &  0.133  & Democrat &   1993-1999 & 0.720 & { 0.947}\\ 
Butler Derrick & 231 & 0.218  & Democrat & 1975-1995 & 1.370 & { 0.925}\\
Richard Ray & 229& 0.239 & Democrat  & 1983-1993 & 1.140 &  { 0.936}\\
Joel Hefley& 182 & 0.143  & Republican & 1987-2007 & 0.753 &  {  0.934}\\
Jon Kyl & 222 & 0.165  & Republican  & 1987-1995&  0.735 &  { 0.936}\\
Carrie Meek & 211 & 0.137 & Democrat &  1993-2003 &   0.735 & {   0.938} \\
Maurice Hinchey& 188 & 0.139 & Democrat & 1993-2013 &   0.731 & {  0.923}\\\hline
    \end{tabular}
    \caption{Information on the selected legislators. The second‑to‑last column reports the relative estimation error of the latent position parameters with respect to the full‑network estimates. The last column reports the AUC obtained by plugging the partial‑information estimates into model~(3) to predict network connections. }
    \label{table:inf_congress_example}
\end{table}
\spacingset{1.9}

Finally, we note that in practice, the computation of the imbalance measure requires full-network latent position estimates, which are typically unavailable under partial information. When node covariates are observed, however,
they can serve as informative proxies. Here 
we use each legislator’s party affiliation to compute the proportion of neighbors from the same party: with two parties in 1990–1994, a balanced neighborhood would yield a proportion near 0.5, whereas strong homophily would push it toward 1.
This proportion is 0.83 for Armey and 0.79 for Furse, indicating that Furse’s neighborhood is more diverse and is consistent with the conclusions drawn from the imbalance measure.

\section{Discussion}\label{sec:discussion}

In this paper, we propose a partial information framework for fitting a general latent space model using a single individual-centered subnetwork. This setting motivates us to develop a new projected gradient descent algorithm and establish its convergence properties. In particular, under conditional likelihood, we characterize how neighborhood structure affects convergence rates, hence providing insight into how to best infer global network properties from localized, partial observations.  Our analysis yields an imbalance measure, primarily intended as a theoretical and diagnostic tool, that links neighborhood structure to estimation accuracy. We validate the theory with simulations and real data applications, {while noting that practical proxies for imbalance may be application-dependent.}

The results have implications for future work on network sampling, particularly in the selection of informative and efficient seeds, and heterogeneous neighborhood effects in peer effect studies.
Taking the U.S. House co-sponsorship network as an example, our results advocate for the collection of information from individuals with local views that exhibit minimal biases, rather than solely focusing on high-impact individuals with high degrees.



The current work focuses on partial information with knowledge depth $L = 2$,   {which should be viewed as one important setting rather than a universally optimal or universally observable choice.} It can serve as a foundation for extending to  $L \geq 2$ in future work. For example, when $L = 3$, the likelihood objective has a similar overall structure to 
$L=2$ and can be optimized using a similar gradient-based algorithm, but several components require modification. Consider the second‑degree neighbor set $\cS' = \{ i \in [n] \mid \exists\, j \in \cS,\ A_{ij} = 1 \}$. Different step sizes (scaled by $|\cS'|$) are needed to update $\alpha_{\cS'}$ and $\alpha_{I-\cS'}$; 
both the centering matrix and the imbalance measure must be redefined for the expanded neighborhood. More generally, as $L$ increases, information expands outward from the center, so neighborhood imbalances persist and continue to affect the accuracy of global parameter estimation.

Aggregating local-view estimates could improve latent parameter estimation, especially in large sparse networks.  Existing divide-and-conquer work for community detection \citep{mukherjee2021two,chakrabarty2025subsampling} does not directly address partial information, and latent space models add rotation and different identifiability constraints across subgraphs.  One promising approach, following \cite{mukherjee2021two}, is to aggregate rotation invariant pairwise quantities  (e.g., distance measures $G$) and weight local views using the imbalance measure (or proxies as in Section~\ref{subsec:congress}).

Another natural extension is to incorporate randomly missing edges, as in \citep{candes2010matrix, abbe2020entrywise}. which may better reflect individuals' perceptions in large-scale social networks, particularly on social media platforms. Finally, the partial information framework itself can be extended to other latent space models, including signed networks \citep{tang2024population}
and longitudinal latent space models \citep{he2024semiparametricmodelinganalysislongitudinal}.

\bibliographystyle{unsrtnat}
\bibliography{Latent}

\end{document}